\documentstyle[12pt,epsfig]{article}
\let\section=\subsection  \let\subsection=\subsubsection

\def\be{\begin{equation}}
\def\ee{\end{equation}}
\def\bea{\begin{eqnarray}}
\def\eea{\end{eqnarray}}

\begin{document}
\begin{center}

{\large \bf Flavored exotic multibaryons and hypernuclei in topological soliton models
}\\[5mm]
V.B. Kopeliovich and A.M. Shunderuk  \\[5mm]
{\small \it
Institute for Nuclear Research, Russian Academy of Sciences, Moscow,
117312 Russia }

\end{center}

\begin{abstract}\noindent
\tenrm \baselineskip=11pt
The energies of baryon states with positive strangeness, or anti-charm (-beauty) are estimated 
in chiral soliton approach, in the "rigid oscillator" version of the bound state soliton model
proposed by Klebanov and Westerberg. Positive strangeness states can appear as relatively 
narrow nuclear
 levels ($\Theta$-hypernuclei), the states with heavy anti-flavors can be bound 
with respect to strong interactions in the original Skyrme variant of the model (SK4 variant).
The binding energies of anti-flavored states are estimated also in the variant of the model with 
6-th order term in chiral derivatives in the lagrangian as solitons stabilizer (SK6 variant). 
The latter variant is less attractive, and nuclear states with anti-charm
and anti-beauty can be unstable relative to strong interactions. The chances to get bound hypernuclei
with heavy anti-flavors increase within "nuclear variant" of the model with rescaled model parameter
(Skyrme constant $e$ or $e'$ decreased by $\sim 30\%$) which is expected to be valid for baryon 
numbers greater than $B\sim 10$. The rational map 
approximation is used to describe multiskyrmions 
with baryon number up to $\sim 30$ and to calculate the quantities necessary for their quantization 
(moments of inertia, sigma-term, etc.).

\end{abstract}
\bigskip
\vspace{0.1cm}

\baselineskip=13pt
\section{Introduction }

The remarkable recent discovery of the positive strangeness pentaquark state
\cite{japan}, its confirmation by several experiments
 \cite{step}
provided strong impact for the searches of other exotic states and revision of existing ideas on
the structure of hadrons and the role of the valence quarks picture in
their description \cite{kl,jw,close,cld,kimm,pqw}.
Subsequently, the discovery of the  strangeness $S=-2$
state with charge $-2$, also manifestly exotic \cite{alt} (see \cite{fw} reviewing previously
existing data), and evidence for a narrow anti-charmed baryon state \cite{akt} have been reported.
Some experiments, however, did not confirm these results, see e.g. \cite{wa89} and \cite{pochod}
where some negative results were summarized and pessimistic point of view was formulated.
The HEP community is waiting now for results of high statistics experiments; some plans for
future pentaquarks searches are presented, e.g. in \cite{burkert}.

The possible existence of such states has been forseen theoretically within the quark
models \cite{j,hs,ro}
\footnote{The parity of lowest exotic states considered here is
negative, see however \cite{kimm}, in difference from the chiral soliton model predictions, where it 
is positive.
 Spin and parity of exotic baryons are not measured yet.},
as well as in chiral soliton models. The prediction of exotic states in chiral soliton models
has not simple and instructive history, from the papers where the exotic antidecuplet and $\{27\}$-
plet of baryons were mentioned \cite{manoh}, a resonant behaviour of the kaon-nucleon phase shift in 
the $\Theta$ channel was obtained in some version of the Skyrme model\cite{km}, first estimates of 
the antidecuplet mass were
 made \cite{bieden, mic}, the masses of exotic baryon states were roughly 
estimated for arbitrary
$B$-numbers \cite{vk}, to papers where more detailed calculations of antidecuplet spectrum were 
performed
 \cite{hans,dpp,herbert}, for recent discussion see also \cite{ja2}. The mass of the dibaryon
with $S=+1,\;I=1/2$ was determined to be only $590\,Mev$ above nucleon-nucleon threshold within soft 
rotator quantization scheme \cite{kss}. It should be
noted that the paper \cite{dpp} which predicted narrow width and low mass of positive strangeness
state called $\Theta^+$
\footnote{As it was admitted recently in \cite{dpp2}, the prediction of the low value of the
mass, $M_\Theta \simeq 1530 Mev$ "was to some extent a luck".} stimulated experimental searches for 
such states, in particular,
 experiments \cite{japan} have been arranged specially to check the 
prediction of \cite{dpp}.

Theoretical ideas and methods which led to the prediction of such states
within the chiral soliton models \cite{hans,dpp,herbert} have been criticized
with quite sound reasoning in \cite{jw} and, in the large $N_c$ limit, in \cite{tc,ikor}.
In the absence of the complete theory of strong interactions
it was in principle not possible to provide the firm predictions for the masses of the states
with accuracy better than about several tens of $Mev$, and similarly for the width of such
states. One can agree with \cite{tc}: the fact that in some cases predictions coincided with
the observed mass of $\Theta^+$ hyperon can be considered as "accidental", see also \cite{dpp2}.

On the other hand, from practical point of view the chiral soliton approach is useful and has 
remarkable predictive power, when at least one of the exotic baryon masses is fitted. The masses of 
exotic 
baryons with strangeness $S=-2$ and isospin $I=3/2$ predicted in this way \cite{wk},
$1.79\;Gev$ for the
 antidecuplet component and $1.85\;Gev$ for the $27$-plet component, are close to 
the value $1.86\;Gev$
 measured later \cite{alt}. Calculations of baryons spectra within chiral soliton approach were made
more recently
 in papers \cite{bfk,mic2,wm1,ekp,dpt}, not in contradiction with \cite{wk}; recent paper 
\cite{w} where the interplay of rotational and vibrational modes
has been investigated, should be mentioned specially. Some reviews and comparison of the chiral 
soliton approach with other models can be found, e.g. in \cite{jm}.

The particular case of strangeness is in certain respect more complicated in comparison with the 
case of other flavors: the rigid rotator
 quantization scheme is not quite valid in this case 
\cite{tc}, whereas the bound state
 approach also is not quite good \cite{ikor}. In the case of 
heavy flavors the rotator quantization
 is not valid at all, but the bound state approach becomes 
more adequite compared
 to strangeness \cite{ikor}.

Baryons with heavy anti-flavors certainly is not a new issue: they have been discussed
in literature long ago,  with various results obtained for the energies of such states.
The strange-anticharmed pentaquark was obtained bound \cite{l} in a quark model with
$(u,d,s)\;SU(3)$ flavor symmetry and in the limit of very heavy $c$-quark.
There were already long ago statements and suggestions in the literature that
 anti-charm or anti-beauty 
can be bound by chiral solitons for the case
 of baryon number $B=1$ \cite{rs,opm} (so called P-baryons).

In \cite{ksr} the mass differences of exotic baryons ($\Theta^+$ and its analogs for anti-charm
and anti-beauty) and nucleons were estimated in the flavor-symmetric limit for decay constants, 
$F_D=F_\pi$ of
 the chiral quark meson model.
In \cite{kz} the anti-flavor excitation energies were calculated in the rigid oscillator version 
\cite{wkleb} of
 the bound state soliton model \cite{ck}, for baryon numbers between $1$ and $8$. 
The rational map
 ansatz for multiskyrmions \cite{hms} has been used as starting configuration in 
the 3-dimensional minimization $SU(3)$ program \cite{kss2}.
These energies were found to be close to $0.59\;Gev$ for anti-strangeness,
$1.75\;Gev$ for anti-charm and $4.95\;Gev$ for anti-beauty, in the latter two
cases these energies are smaller than masses of $D$ and $B$-mesons which
enter the lagrangian \cite{kz}. The flavor symmetry breaking in flavor decay constants 
($F_D/F_\pi >1$) plays important role for these estimates. So, it was clear hint that such baryonic 
systems can be
 bound relative to strong interactions.

Similar results, in principle, follow from recent analysis within bound state soliton model 
\cite{ikor},
 and also within the diquark model \cite{jw}.
 The spectra of exotic states with heavy 
flavors have been estimated in different models, already after discovery of positive strangeness 
pentaquark \cite{vkj} (any baryon number), \cite{cheung,hldcz,bic,ch,kimm2} and others. The 
possibility of existence of nuclear matter fragments with positive strangennes was discussed 
recently in \cite{mill}.

In present paper we estimate the energies of ground states of multibaryons with baryon numbers up
to $\sim 30$ with different (anti)flavors using a very transparent "rigid oscillator" model \cite{wkleb}.
In the next section the properties of multiskyrmions are considered which are necessary to
calculate the energies of flavor excitations, using the
 rational map approximation for $B>1$\cite{hms}.
It is shown that  $\Theta^+$ baryon is bound by nuclear systems, providing positive 
strangeness multibaryons ($\Theta$-hypernuclei), their binding energy can reach several tens of $Mev$.
The multiskyrmion configurations possess some remarkable scaling properties, as a
result, the flavor and antiflavor excitation energies are close
 to those for $B=1$. 
The quantization scheme (slightly modified rigid oscillator version \cite{wkleb}) is described
in section 3 where the flavor and antiflavor excitation energies are calculated
as well. The masses (binding energies) of ground states of positive strangeness states
 - 
$\Theta$-hypernuclei - are presented in section 4, followed by those for anti-charmed or -beautiful 
states. The last section contains some conclusions and prospects.
\section{Properties of multiskyrmions}
Here we calculate the properties of multiskyrmion configurations necessary for
calculation of flavor excitation energies and hyperfine splitting constants
which govern the $1/N_c$ corrections to the energies of the quantized states.
As it was already noted previously, the details of baryon-baryon interactions
do not enter the calculations explicitly, although they influence implicitly
 via
the integral characteristics of the bound states of skyrmions shown in {\bf Tables 1,2}.

The lagrangian of the Skyrme model in its well known form depends on parameters 
$F_{\pi}, \; F_D$ and $e$ and
 can be written in the following way \cite{skyrme,witten}:
\bea
\label{L}
{\cal L} & = & - \frac{F_\pi^2}{16}Tr l_{\mu} l^{\mu} + {1 \over 32e^2}
Tr [l_\mu,l_\nu]^2 +\frac{F_\pi^2m_\pi^2}{16} Tr(U+U^{\dagger}-2) + \nonumber \\
 & + &\frac{F_D^2m_D^2-F_\pi^2m_\pi^2}{24}Tr(1-\sqrt{3}\lambda_8)(U+U^{\dagger}-2) +\nonumber \\
& + & \frac{F_D^2-F_\pi^2}{48} Tr(1-\sqrt{3}\lambda_8)(Ul_\mu l^\mu +
l_\mu l^\mu U^{\dagger}).
\eea
Here $U \in SU(3)$ is a unitary matrix incorporating chiral (meson) fields, and
$l_\mu= \partial _\mu U U^\dagger$. In this model $F_\pi$ is fixed
at the physical value: $F_\pi$ = $186\, Mev$ and $m_D$ is the mass of $K, \, D$ or
$B$ meson. The ratios $F_D/ F_\pi$ are known to be $1.22$ and $2.28^{+1.4}_{-1.1}$ for, respectively,
kaons and $D$-mesons.
 The Skyrme parameter $e$ is close to $4$ in numerical fits of the hyperons 
spectra (see discussion at the end of this section).
In the variant of the model with 6-th order term as solitons stabilizer 
the contribution is added to the lagrangian density \cite{jack,lm,fp}
\be
\label{L_6}
L_6 = - {c_6\over 48} Tr [l_\mu,l^\nu] [l_\nu,l^\alpha] [l_\alpha,l^\mu],
\ee
where we introduced the coefficient $1/48$ in the definition of the constant
$c_6$ for further convenience.
 It is known that this term can be considered as approximation to 
the exchange of $\omega$-meson in the limit $m_\omega \to \infty$ \cite{jack} \footnote{We are using 
in (\ref{L_6}) one of several possible forms of 6-th order term, but all give the same contribution to
the static mass of the $SU(2)$ solitons, see also discussion in \cite{jack}. General consideration of 
high order terms and their role for skyrmion properties can be found in \cite{lm}.}.
The flavor symmetry breaking $(FSB)$ in the lagrangian is of
the usual form, and is sufficient to describe the mass splittings of the octet and decuplet of
baryons within the collective coordinate quantization approach \cite{sw}.
A nice and useful feature of the lagrangian (\ref{L},\ref{L_6}) is that it contains only second
power of the time derivative, and this allows to perform quantization without problems (next section).

The Wess-Zumino term, to be added to the action, which can be written as a 5-dimensional
differential form \cite{witten}, plays an
 important role in the quantization procedure:
\be
\label{WZ}
 S^{WZ}= \frac{-iN_c}{240 \pi^2}\int_\Omega d^5x\epsilon^{\mu\nu\lambda\rho
\sigma} Tr (l_\mu l_\nu l_\lambda l_\rho l_\sigma), \ee
where $\Omega $ is a 5-dimensional region with the 4-dimensional space-time
Action (\ref{WZ})  defines important topological properties of skyrmions, but it
does not contribute to the static masses of classical configurations  \cite{g,mic}.
Variation of this action can be presented as a well defined contribution
to the lagrangian (integral over the 4-dimensional space-time).

We begin our calculations with $U \in SU(2)$.
The classical mass of $SU(2)$ solitons, in the most general case, depends on $3$
profile functions: $f, \, \alpha$ and $\beta$, and is given by
\be
\label{mcl}
M_{cl} = \int \biggl\{ \frac{F_\pi^2}{8} \bigl[\vec{l}_1^2+\vec{l}_2^2+
\vec{l}_3^2 \bigr] + {1 \over 2e^2} \bigl[ [\vec{l}_1\vec{l}_2]^2 +
[\vec{l}_2\vec{l}_3]^2 +[\vec{l}_3\vec{l}_1]^2 \bigr] +
 {1 \over 4} F_\pi^2m_\pi^2 (1-c_f) + 2 c_6 (\vec{l}_1 \vec{l}_2 \vec{l}_3)^2
 \biggr\} d^3r .
\ee
Here $\vec{l}_k$ are the $SU(2)$ chiral derivatives defined by
$ \vec{\partial}UU^\dagger = i \vec{l}_k \tau_k$, where $k=1,2,3$.
The general parametrization of $U_0$ for an $SU(2)$ soliton we use here
is given by
$U_0 = c_f+s_f \vec{\tau}\vec{n}$ with $n_z=c_{\alpha}$, $n_x=s_{\alpha}
c_{\beta}$, $n_y=s_{\alpha}s_{\beta}$, $s_f=sinf$, $c_f=cosf$, etc.

For the rational map ($RM$) ansatz we are using here as starting configurations \cite{hms},
\be
n_x=\frac{2 Re\, R(\xi)}{1+|R(\xi)|^2}, \qquad
n_y=\frac{2 Im\, R(\xi)}{1+|R(\xi)|^2}, \qquad
n_z=\frac{1-|R(\xi)|^2}{1+|R(\xi)|^2}, \ee
where $R(\xi)$ is a ratio of polynomials of the maximal power $B$ of the variable
$\xi=tg(\theta /2)exp(i\phi)$, $\theta$ and $\phi$ being polar and azimuthal
angles defining the direction of the radius-vector $\vec{r}$. It is important assumption
that vector $\vec{n}$ depends on angular variables, but does not depend on $r$, whereas 
the profile $f(r)$ depends on distance from the soliton center, only. The explicit
form of $R(\xi)$ is given in \cite{hms,bps} for different values of $B$. 
Within $RM$ approximation all characteristics of multiskyrmions necessary for us
(including mass and moments of inertia) depend on two quantities, integrals over angular variables,
\bea
\label{NI}
{\cal N} = {1\over 8\pi} \int r^2 (\partial_i n_k)^2 d\Omega, \qquad 
{\cal I} = {1\over 8\pi} \int r^4 [\vec{\partial}n_i \vec{\partial}n_k]^2 d\Omega,
\eea
satisfying inequality ${\cal I} \geq {\cal N}^2$ \cite{hms}. For configuration of lowest energy
${\cal N} = B$, $f(0)-f(\infty) =\pi$ and the value ${\cal I}$ should be found by minimization
of the map $S^2 \to S^2$ \cite{hms}.
The classical mass of multiskyrmion then simplifies to:
\be
\label{mass}
M_{cl}= 4\pi \int \Biggl[{F_\pi^2\over 8}\Biggl(f'^2 + 2B{s_f^2\over r^2}\Biggr) +
{s_f^2\over 2e^2r^2}\Biggl(2f'^2B+s_f^2{{\cal I}\over r^2}\Biggr)+ 4 c_6{\cal I}f'^2{s_f^4\over r^4} +
\rho_{M.t.}\Biggr] r^2dr,
\ee
which should and can be minimized easily for definite $B$ and ${\cal I}$. The mass term density
is simple for starting $SU(2)$ skyrmion: $\rho_{M.t.}= F_\pi^2m_\pi^2(1-c_f)/4$.
The quantity $\lambda$ can be introduced \cite{fp} which characterizes the relative weight of
the 6-th order term, according to $\lambda/(1-\lambda )^2 = c_6 F_\pi^2 e^4$, or $c_6 = \lambda/
(F_\pi^2 e'^4)$. For pure SK6 variant ($\lambda=1,\; e\to \infty$ and $e'=e(1-\lambda)$-fixed),
there is relation $c_6=1/(F_\pi^2 e'^4)$.

The ``flavor" moment of inertia plays a very important role in the procedure
of $SU(3)$ quantization \cite{g,hans}, see formulas (\ref{LQ},\ref{H},\ref{om}) etc. below.
It defines the $SU(3)$ rotational energy 
$E_{rot}(SU_3)=\Theta_F(\Omega_4^2+\Omega_5^2+\Omega_6^2+\Omega_7^2)/2$ with
$\Omega_a$, $a=4,...7$ being the angular velocities of rotation in $SU(3)$ configuration space.
For $SU(2)$ skyrmions as starting configurations and the $RM$ ansatz describing classical field
configurations $\Theta_F$ is given by \cite{vk1,ash}:
\bea
\label{tf}
 \Theta_F  = {1 \over 8} \int (1-c_f)\biggl[ F_D^2 + {1 \over e^2}
 \biggl( f'^2 + 2B {s_f^2 \over r^2} \biggr) 
  + 2 c_6 {s_f^2\over r^2}\biggl(2Bf'^2 + {\cal I}{s_f^2\over r^2} \biggr) \biggr] r^2dr .
 \eea
It is simply connected with $\Theta_F^{(0)}$ of the flavor symmetric case
($F_D=F_\pi$):
\be\Theta_F=\Theta_F^{(0)}+(F_D^2/F_\pi^2-1) \Gamma/4, \ee
with $\Gamma$  defined in (\ref{G}) below.
\begin{center}
\begin{tabular}{|l|l|l|l|l|l|l|l|l|}
\hline
$B$ & $\Theta_I^{SK4}$ & $\Theta_F^{(0)SK4}$ & $\Gamma^{SK4}$ & $\tilde{\Gamma}^{SK4}$ &
$\Theta_I^{SK6}$ & $\Theta_F^{(0)SK6}$ & $\Gamma^{SK6}$ & $\tilde{\Gamma}^{SK6}$ \\ \hline
$1 $ & $5.56$ & $2.05$ & $4.80$ & $14.9$ & $5.13$ & $2.28$ & $6.08$ & $15.8$ \\ \hline
$2 $ & $11.5$ & $4.18$ & $9.35$ & $22.0$ & $9.26$ & $4.94$ & $14.0$ & $24.7$ \\ \hline
$3 $ & $14.4$ & $6.34$ & $14.0$ & $27.0$ & $12.7$ & $7.35$ & $20.7$ & $30.4$ \\ \hline
$4 $ & $16.8$ & $8.27$ & $18.0$ & $31.0$ & $15.2$ & $8.93$ & $24.5$ & $33.7$ \\ \hline
$5 $ & $23.5$ & $10.8$ & $23.8$ & $35.0$ & $18.7$ & $11.8$ & $32.8$ & $38.3$ \\ \hline
$6 $ & $25.4$ & $13.1$ & $29.0$ & $38.0$ & $21.7$ & $14.1$ & $39.3$ & $41.6$ \\ \hline
$7 $ & $28.9$ & $14.7$ & $32.3$ & $44.0$ & $23.9$ & $15.4$ & $42.5$ & $43.4$ \\ \hline
$8 $ & $33.4$ & $17.4$ & $38.9$ & $47.0$ & $27.2$ & $18.5$ & $51.6$ & $46.9$ \\ \hline
$9 $ & $37.8$ & $20.6$ & $46.3$ & $47.5$ & $30.2$ & $21.1$ & $59.1$ & $49.7$ \\ \hline
$10$ & $41.4$ & $23.0$ & $52.0$ & $50.0$ & $32.9$ & $23.5$ & $65.8$ & $51.9$ \\ \hline
$11$ & $45.2$ & $25.6$ & $58.5$ & $52.4$ & $35.8$ & $26.1$ & $73.6$ & $54.3$ \\ \hline
$12$ & $48.5$ & $28.0$ & $64.1$ & $54.6$ & $38.4$ & $28.3$ & $79.9$ & $56.2$ \\ \hline
$13$ & $52.1$ & $30.5$ & $70.2$ & $56.8$ & $41.2$ & $30.8$ & $87.1$ & $58.1$ \\ \hline
$14$ & $56.1$ & $33.6$ & $78.2$ & $58.9$ & $44.3$ & $34.0$ & $96.9$ & $60.5$ \\ \hline
$15$ & $59.8$ & $36.3$ & $85.1$ & $60.9$ & $47.1$ & $36.7$ & $105 $ & $62.4$ \\ \hline
$16$ & $63.2$ & $38.9$ & $91.5$ & $62.8$ & $49.7$ & $39.3$ & $112 $ & $64.1$ \\ \hline
$17$ & $66.2$ & $41.2$ & $96.8$ & $64.6$ & $52.1$ & $41.3$ & $118 $ & $65.4$ \\ \hline
$18$ & $70.3$ & $44.5$ & $106 $ & $66.4$ & $55.2$ & $44.8$ & $129 $ & $67.5$ \\ \hline
$19$ & $73.9$ & $47.4$ & $113 $ & $68.2$ & $58.0$ & $47.8$ & $138 $ & $69.2$ \\ \hline
$20$ & $77.5$ & $50.4$ & $121 $ & $69.9$ & $60.8$ & $50.8$ & $147 $ & $70.8$ \\ \hline
$21$ & $80.9$ & $53.2$ & $128 $ & $71.5$ & $63.5$ & $53.6$ & $156 $ & $72.4$ \\ \hline
$22$ & $84.3$ & $56.0$ & $136 $ & $73.1$ & $66.1$ & $56.4$ & $164 $ & $73.8$ \\ \hline
$23$ & $88.0$ & $59.2$ & $144 $ & $74.7$ & $69.0$ & $59.7$ & $174 $ & $75.4$ \\ \hline
$24$ & $91.3$ & $62.0$ & $151 $ & $76.2$ & $71.6$ & $62.5$ & $183 $ & $76.7$ \\ \hline
$25$ & $94.7$ & $64.9$ & $159 $ & $77.6$ & $74.2$ & $65.4$ & $192 $ & $78.0$ \\ \hline
$26$ & $98.2$ & $68.1$ & $168 $ & $79.1$ & $77.0$ & $68.7$ & $202 $ & $79.4$ \\ \hline
$27$ & $102 $ & $71.1$ & $176 $ & $80.5$ & $79.7$ & $71.7$ & $211 $ & $80.8$ \\ \hline
$28$ & $105 $ & $74.3$ & $185 $ & $81.9$ & $82.5$ & $75.1$ & $222 $ & $82.2$ \\ \hline
$32$ & $118 $ & $86.4$ & $217 $ & $87.2$ & $93.0$ & $87.4$ & $260 $ & $86.9$ \\ \hline
\end{tabular}
\end{center}

{\bf Table 1.} {\small Static characteristics of multiskyrmions - moments of inertia and $\sigma$-term
$\Gamma$, $\tilde{\Gamma}$ in the $SK4$ variant of the model with $e=4.12$, and for the $SK6$
variant of the model, $e'=4.11$, in $Gev^{-1}$.}\\

The isotopic momenta of inertia are the components of the corresponding
tensor of inertia presented and discussed in many papers, see e.g.
\cite{g,hans,vk1}. For majority of multiskyrmions we discuss, this tensor of inertia is 
close to the unit matrix multiplied by the isotopic moment of inertia: $\Theta_{ab} \simeq \Theta_I
\delta_{ab}$,  $\Theta_I=\Theta_{I,aa}/3$.
This is exactly the case for $B=1$ and, to within
a good accuracy, for $B=3,\,7 $. Considerable deviations take place for the
torus with $B=2$, smaller ones for $B=4,\,5, \,6$, and generally, they decrease with increasing
$B$-number. We shall use for our estimates very simple expression obtained within rational
map approximation \cite{vk1,ash}
\be
\label{ti}
\Theta_I = {4\pi\over 3}\int s_f^2\biggl[{F_\pi^2\over 2}+{2\over e^2}\biggl(f'^2 + 
 B{s_f^2\over r^2}\biggr) + 8 c_6 B s_f^4 {f'^2\over r^2} \biggr] r^2dr
\ee
The isotopic inertia (\ref{ti}) at large enough baryon numbers receives the main
contribution from the spherical envelope of multiskyrmion where its mass is concentrated. The 
dimensions of this spherical bubble grow like $R_B \sim \sqrt{B}$ \cite{vk1}, and moments of inertia
are proportional roughly to the baryon number.

The quantity $\Gamma$ (or $\Sigma$-term) defines the contribution of the
mass term to the classical mass of solitons,  $\tilde{\Gamma}$ enters due to the presence of
$FSB$ term proportional to the difference $F_D^2 -F_\pi^2$
 in (\ref{L}), the last term in (\ref{L}).
They define the potential where the rigid oscillator moves, and are given by:
\be 
\label{G}
\Gamma = \frac{F_{\pi}^2}{2} \int (1-c_f) d^3\vec{r}, \qquad
  \tilde{\Gamma}={1 \over 4} \int c_f
 \bigl[ (\vec{\partial}f)^2 +s_f^2(\vec{\partial}n_i)^2\bigr] . \ee
The following relation can also be established:
$\tilde{\Gamma}=2(M_{cl}^{(2)}/F_\pi^2-e^2\Theta_F^{SK4})$, where
$M_{cl}^{(2)}$
is the second-order term contribution to the classical mass of the soliton,
and $\Theta_F^{SK4}$ is the Skyrme term contribution to the flavor moment of
 inertia.
The calculated momenta of inertia $\Theta_F$, $\Theta_I$,
$\Gamma$ or $\Sigma$-term and $\tilde{\Gamma}$ for solitons with baryon numbers up to 32
are presented in {\bf Tables 1, 2}.

 Sigma-term $\Gamma$ gets contribution from the whole volume of
multiskyrmion, where $c_f\sim -1$, and by this reason it grows faster than moment of inertia 
$\Theta_I$. Flavor inertia $\Theta_F$ gets contribution from the surface and the volume of 
multiskyrmion, and its behaviour is intermediate between that of $\Gamma$ and $\Theta_I$.

For both variants of the model, $SK4$ and $SK6$, we calculated static characteristics of 
multiskyrmions
for two values of the only parameter of the model, constant $e$, or $e'$ for the $SK6$ variant
(the connection between coefficient $c_6$ and $e'$ is $e'= 1/(F_\pi^2 c_6)^{1/4}$).
For the $SK4$ variant of the model and $e=4.12$ the numbers given in {\bf Table 1} for $B=1\;-\;8$ are
obtained as a result of direct numerical energy minimization in $3$ dimensions performed using the
calculation algorythm developed in \cite{kss2}. By this reason they
differ slightly from those obtained in pure rational map approximation.
This difference is the largest for the case of $B=2$ and decreases with increasing $B$.
In all other cases we used the $RM$ approximation with values of the Morse function ${\cal I}$
given in \cite{hms,bps}.

The second value of the constants, $e=3.00$ and $e'=2.84$, leads to the "nuclear variant" of the model 
which allowed to
 describe quite successful the mass splittings of nuclear isotopes for atomic 
(baryon) numbers between
 $\sim 10$ and $\sim 30$ \cite{ksm}.
The static characteristics of multiskyrmions change considerably when the change of the constants $e$
or $e'$ is made by about $30\%$, see {\bf Table 2}, since dimensions of solitons scale 
like $1/(F_\pi e)$, and the isotopic mass splittings scale like $F_\pi e^3$. However. the flavor 
excitation energies change not crucially, even slightly
 for charm and beauty, according to the scale 
invariance of these quantities \cite{vk1}, as described in the next section.
\begin{center}
\begin{tabular}{|l|l|l|l|l|l|l|l|l|}
\hline
$B$ & $\Theta_I^{SK4^*}$ & $\Theta_F^{(0)SK4^*}$ & $\Gamma^{SK4^*}$ & $\tilde{\Gamma}^{SK4^*}$ &
 $\Theta_I^{SK6^*}$ & $\Theta_F^{(0)SK6^*}$ & $\Gamma^{SK6^*}$ & $\tilde{\Gamma}^{SK6^*}$ \\ \hline
$1 $ & $12.8$ & $4.66$ & $10.1$ & $19.6$ & $14.2$ & $6.21$ & $15.3$ & $22.3$ \\ \hline
$2 $ & $24.3$ & $9.87$ & $20.9$ & $28.8$ & $25.7$ & $13.6$ & $35.9$ & $34.7$ \\ \hline
$3 $ & $34.7$ & $15.1$ & $31.7$ & $35.6$ & $35.5$ & $20.4$ & $53.9$ & $42.5$ \\ \hline
$4 $ & $42.9$ & $19.4$ & $40.1$ & $41.1$ & $43.2$ & $25.0$ & $64.6$ & $46.9$ \\ \hline
$5 $ & $53.5$ & $25.4$ & $53.2$ & $46.2$ & $52.9$ & $32.9$ & $86.2$ & $53.1$ \\ \hline
$6 $ & $62.6$ & $30.7$ & $64.7$ & $50.6$ & $61.4$ & $39.4$ & $103 $ & $57.4$ \\ \hline
$7 $ & $69.6$ & $34.9$ & $72.5$ & $54.4$ & $68.0$ & $43.3$ & $112 $ & $59.8$ \\ \hline
$8 $ & $79.9$ & $41.3$ & $87.4$ & $58.2$ & $77.3$ & $51.7$ & $135 $ & $64.4$ \\ \hline
$9 $ & $88.9$ & $47.1$ & $101 $ & $61.7$ & $85.7$ & $58.9$ & $154 $ & $67.9$ \\ \hline
$10$ & $97.4$ & $52.6$ & $113 $ & $64.9$ & $93.5$ & $65.3$ & $171 $ & $70.8$ \\ \hline
$11$ & $106 $ & $58.5$ & $126 $ & $67.9$ & $102 $ & $72.5$ & $191 $ & $73.8$ \\ \hline
$12$ & $114 $ & $63.8$ & $138 $ & $70.8$ & $109 $ & $78.7$ & $207 $ & $76.1$ \\ \hline
$13$ & $122 $ & $69.5$ & $151 $ & $73.6$ & $117 $ & $85.4$ & $225 $ & $78.6$ \\ \hline
$14$ & $132 $ & $76.3$ & $168 $ & $76.3$ & $125 $ & $94.0$ & $249 $ & $81.5$ \\ \hline
$15$ & $140 $ & $82.3$ & $182 $ & $78.8$ & $133 $ & $101 $ & $269 $ & $83.9$ \\ \hline
$16$ & $148 $ & $88.1$ & $196 $ & $81.2$ & $141 $ & $108 $ & $287 $ & $86.0$ \\ \hline
$17$ & $155 $ & $93.2$ & $207 $ & $83.5$ & $148 $ & $114 $ & $302 $ & $87.6$ \\ \hline
$18$ & $164 $ & $100 $ & $225 $ & $85.9$ & $156 $ & $123 $ & $328 $ & $90.1$ \\ \hline
$19$ & $173 $ & $107 $ & $241 $ & $88.1$ & $164 $ & $131 $ & $350 $ & $92.2$ \\ \hline
$20$ & $181 $ & $113 $ & $257 $ & $90.3$ & $172 $ & $139 $ & $372 $ & $94.1$ \\ \hline
$24$ & $213 $ & $138 $ & $320 $ & $98.2$ & $202 $ & $170 $ & $457 $ & $101 $ \\ \hline
$28$ & $245 $ & $165 $ & $387 $ & $105 $ & $232 $ & $202 $ & $550 $ & $107 $ \\ \hline
$32$ & $275 $ & $191 $ & $454 $ & $112 $ & $261 $ & $234 $ & $640 $ & $113 $ \\ \hline
\end{tabular}
\end{center}

{\bf Table 2.} {\small Static characteristics of multiskyrmions - moments of inertia and $\Gamma$,
$\tilde{\Gamma}$ for rescaled, or nuclear variants of the model: $e=3.00$ in the $SK4$ 
and
 $e'=2.84$ for the $SK6$ variants, also in $Gev^{-1}$.}\\

\section{Flavor and antiflavor excitation energies}

The $SU(3)$ effective action
 defined by (\ref{L},\ref{WZ}) leads to the collective lagrangian 
obtained in \cite{g}.
 To quantize the solitons in their $SU(3)$ configuration space,  in the
spirit of the bound state approach to the description of strangeness
proposed in \cite{ck,wkleb} and used in \cite{vk1,kz},
 we consider the collective coordinate 
motion of the
 meson fields incorporated into the matrix $U$:
\be 
\label{AS}
U(r,t) = R(t) U_0(O(t)\vec{r}) R^{\dagger}(t), \qquad
 R(t) = A(t) S(t), \ee
where $U_0$ is the $SU(2)$ soliton embedded into $SU(3)$ in the usual way (into
the upper left hand  corner), $A(t) \in SU(2)$ describes $SU(2)$ rotations and
$S(t) \in SU(3)$ describes
rotations in the ``strange", ``charm" or ``beauty" directions
and $O(t)$ describes rigid rotations in real space.
In the quantization procedure of the rotator with the help of $SU(3)$ collective coordinates
the following definition of angular velocities in $SU(3)$ configuration space is
accepted \cite{g}:
\be \label{Omega}
R^\dagger(t)\dot{R}(t) = - {i\over 2} \Omega_\alpha \lambda_\alpha,
\ee
with $\alpha=1,...,8$, $\lambda_\alpha$ being $SU(3)$ Gell-Mann matrices. For the quantization
method proposed in \cite{wkleb} and used here the parametrization (\ref{AS}) is more convenient,
the components $\Omega_\alpha$ can be expressed via collective coordinates introduced in (\ref{AS}).

For definiteness we consider the extension of the $(u,d)$ $SU(2)$
Skyrme model in the $(u,d,s)$ direction, when $D$ is the field of $K$-mesons
,
but it is clear that quite similar extensions can also be made
in the directions of charm or bottom. So
\be S(t) = exp (i  {\cal D} (t)),  \qquad
 {\cal D} (t) = \sum_{a=4,...7} D_a(t) \lambda_a, \ee
where $\lambda_a$ are Gell-Mann matrices of the $(u,d,s)$, $(u,d,c)$ or $(u,d,b)$
$SU(3)$ groups. The $(u,d,c)$ and $(u,d,b)$ $SU(3)$
groups are quite analogous to
the $(u,d,s)$ one. For the $(u,d,c)$ group a simple redefiniton of
hypercharge should be made. For the $(u,d,s)$ group,
 $D_4=(K^++ K^-)/\sqrt{2}$, $D_5=i(K^+-K^-)/\sqrt{2}$, etc.
For the $(u,d,c)$ group $D_4=(D^0 + \bar{D}^0)/\sqrt{2}$, etc.

The angular velocities of the isospin rotations $\vec{\omega}$ are defined
in the standard way \cite{g}:
$ A^{\dagger} \dot{A} =-i \vec{\omega} \vec{\tau}/2. $
We shall not consider here the usual space rotations in detail because the
corresponding momenta of inertia for baryonic systems ($BS$) are much greater than the
isospin
 momenta of inertia, and for the lowest possible values of angular momentum
$J$, the
 corresponding quantum correction is either exactly
zero (for even $B$), or small.

The field $D$ is small in magnitude. In fact, it is, at least,
 of order $1/\sqrt{N_c}$, where $N_c$ is the number of colors in $QCD$,
 see Eq. (\ref{d}).
Therefore, the expansion of the matrix $S$ in $D$ can be made safely.

The mass term of the lagrangian (\ref{L}) can be calculated exactly, without
expansion in the powers of the field $D$, because the matrix $S$ is given by
 \cite{wkleb}
 $ S=1+i{\cal D} \; sin\,d/ d
-{\cal D}^2 (1-cos\,d)/ d^2 $ with $Tr {\cal D}^2=2d^2 $.
We find that
\be \Delta{\cal L}_M=-\frac{F_D^2m_D^2-F_\pi^2m_\pi^2}{4} (1-c_f)s_d^2 \ee
The expansion of this term can be done easily up to any order in $d$.
The comparison of this expression with $\Delta L_M$, within the collective
coordinate approach of the quantization of $SU(2)$ solitons in the $SU(3)$
configuration space \cite{g,hans}, allows us to establish
 the relation $sin^2\, d =sin^2 \nu$,
where $\nu$ is the angle of the $\lambda_4$ rotation, or the rotation into
the ``strange" (``charm", ``beauty" ) direction.
After some calculations we find that the Lagrangian
of the model, to the lowest order in the field $D$, can be written as
\bea 
\label{LQ}
 L &=&-M_{cl,B}+4\Theta_{F,B} \dot{D}^{\dagger}\dot{D}-\biggl[\Gamma_B \biggl(
\frac{F_D^2}{F_{\pi}^2} m_D^2-m_{\pi}^2 \biggr)+ \tilde{\Gamma}_B(F_D^2-F_\pi^2)
\biggr] D^{\dagger}D - \nonumber \\
&-&
 i{N_cB \over 2}(D^{\dagger}\dot{D}-\dot{D}^{\dagger}D).
\eea
Here and below $D$ is the doublet $K^+, \, K^0$ ($D^0, \,
D^-$, or $B^+,\,B^0$): $d^2=Tr {\cal D}^2/2=2D^\dagger D$.
We have kept the standard notation for the moment of inertia of the
rotation into the ``flavor" direction $\Theta_F$ for $\Theta_c, \,
\Theta_b$ or $\Theta_s$ \cite{g,sw}; different notations are used in
\cite{wkleb} (the index $c$ denotes the
charm quantum number, except in $N_c$). The contribution proportional to
$\tilde{\Gamma}_B$ is suppressed in comparison with the term $\sim \Gamma$
by a small factor $\sim (F_D^2-F_\pi^2)/m_D^2$, and is more
important for strangeness.

The term proportional to $N_cB$ in (\ref{L}) arises from the Wess-Zumino term
in the action and is responsible for the difference of
the excitation energies of strangeness and antistrangeness
(flavor and antiflavor in the general case) \cite{ck,wkleb}.

Following the canonical quantization procedure the Hamiltonian of the
system, including the terms of the order
of $N_c^0$, takes the form \cite{wkleb}:
\be
\label{H}
H_B=M_{cl,B} + {1 \over 4\Theta_{F,B}} \Pi^{\dagger}\Pi +
\biggl[\Gamma_B
\bar{m}^2_D+\tilde{\Gamma}_B(F_D^2-F_\pi^2)+\frac{N_c^2B^2}{16\Theta_{F,B}}
\biggr] D^{\dagger}D +i {N_cB \over 8\Theta_{F,B}}
(D^{\dagger} \Pi- \Pi^{\dagger} D), \ee
where $\bar{m}_D^2 = (F_D^2/F_{\pi}^2) m_D^2-m_\pi^2$.
The momentum $\Pi$ is canonically conjugate to variable $D$ (see Eq.$(18)$
below).
Eq. (\ref{H}) describes an oscillator-type motion of the field $D$
 in the background formed
by the $(u,d)$ $SU(2)$ soliton. After the diagonalization, which can be done
explicitly following \cite{wkleb}, the normal-ordered Hamiltonian can be
written as
\be H_B= M_{cl,B} + \omega_{F,B} a^{\dagger} a + \bar{\omega}_{F,B} b^{\dagger} b
 + O(1/N_c) \ee
with $a^\dagger$, $b^\dagger$ being the operators of creation of strangeness
(i.e., antikaons) and antistrangeness
(flavor and antiflavor) quantum number, $\omega_{F,B}$ and
$\bar{\omega}_{F,B}$ being the
frequencies of flavor (antiflavor) excitations. $D$ and $\Pi$ are connected
with $a$ and $b$ in the following way \cite{wkleb}:
\be D^i= \frac{1}{\sqrt{N_cB\mu_{F,B}}}(b^i+a^{\dagger i}), \qquad
\Pi^i = \frac{\sqrt{N_cB\mu_{F,B}}}{2i}(b^i - a^{\dagger i}) \ee
with
\be
\label{mu}
\mu_{F,B} =\bigl[ 1 + 16 [\bar{m}_D^2 \Gamma_B+(F_D^2-F_\pi^2)\tilde{\Gamma}_B]
 \Theta_{F,B}/ (N_cB)^2 \bigr]^{1/2}. \ee
$\mu_{F,B}$ is slowly varying quantity, it simplifies for large mass $m_D$:
\be
\label{mu1}
\mu_{F,B} \to 4\bar{m}_D \frac{\sqrt{\Gamma_B \Theta_{F,B}}}{N_cB}.
 \ee
Obviously, at large $N_c$,  $\mu \sim N_c^0\sim 1$, and dependence on the $B$-number is also
weak, since both $\Gamma_B, \;\Theta_{F,B} \sim\,N_cB$ \footnote{Strictly, at large $B$, $\Gamma_B
\sim B^{3/2}$ as explained above. But numerically at $B<30$,  $\Gamma_B\sim B$, as can be seen from 
Tables 1 and 2.}.
For the lowest states the values of $D$ are small:
\be
\label{d}
|D| \sim\bigl[16\Gamma_B\Theta_{F,B}\bar{m}_D^2+N_c^2B^2 \bigr]^{-1/4},\ee
and increase, with increasing flavor number $|F|$, like $(2|F|+1)^{1/2}$.
As it follows from (\ref{d}) \cite{wkleb,kz}, deviations of the field $D$ from the vacuum
decrease with increasing mass $m_D$, as well as with increasing number of
colors $N_c$, and this explains why the method works for any $m_D$, including charm and 
beauty quantum numbers.

The excitation frequencies $\omega$ and $\bar{\omega}$ are:
\be
\label{om}
 \omega_{F,B} = \frac{N_cB}{8\Theta_{F,B}} ( \mu_{F,B} -1 ), \qquad
 \bar{\omega}_{F,B} = \frac{N_cB}{8\Theta_{F,B}} ( \mu_{F,B} +1 ) .\ee
 The oscillation time can be estimated as $\tau_{osc}\sim \pi /\omega_{F,B} \sim
 2\pi (\Theta_B/\Gamma_B)^{1/2}/m_D$, so, it decreases with increasing $m_D$.
As  was observed in \cite{vk1,kz}, the difference
$\bar{\omega}_{F,B}-\omega_{F,B}=N_cB/(4\Theta_{F,B})$ coincides, to the
leading order in $N_c$, with the expression obtained in the collective coordinate
 approach \cite{g,sw}, see Appendix.  At large $m_D$ using (\ref{mu1})
for the difference $\omega_{F,1}-\omega_{F,B}$ we obtain $(N_c=3)$:
\be
\label{omdif}
\bar{\omega}_{F,1}-\bar{\omega}_{F,B} \simeq \frac{\bar{m}_D}{2} \biggl[\biggl
(\frac{\Gamma_1}
{\Theta_{F,1}}\biggr)^{1/2}-\biggl(\frac{\Gamma_B}{\Theta_{F,B}}
\biggr)^{1/2}\biggr] +{3 \over 8}\biggl( {B \over \Theta_{F,B}}-
 {1 \over \Theta_{F,1}} \biggr).\ee
Obviously, at large $m_D$, the first term in (\ref{omdif}) dominates and is
positive if $\Gamma_1/
 \Theta_{F,1} \geq \Gamma_B/ \Theta_{F,B}$. This is confirmed by looking at
{\bf Table 1}. Note also that the bracket in the first term in (\ref{omdif}) does
not depend on the parameters of the model if the background $SU(2)$
soliton is calculated in the chirally symmetrical limit: both $\Gamma$ and
$\Theta$
 scale like $ \sim 1/(F_\pi e^3)$. In a realistic case when the physical pion
mass is included in (\ref{L}) there is some weak dependence on the parameters of the model.

The $FSB$ in the flavor decay constants, i.e. the fact that $F_K/F_\pi
\simeq 1.22$ and $F_D/F_\pi=2.28^{+1.4}_{-1.1}$, should be taken into account.
 In the Skyrme model this fact leads to the increase of the flavor excitation
frequencies which changes the spectra of flavored $(c,\,b)$ baryons and
puts them in a better agreement
 with the data \cite{rs}. It also leads
 to some changes of the total binding energies of $BS$
\cite{kz}. This is partly due to the large contribution of the Skyrme term
to the flavor moment of inertia $\Theta_F$. Note, that in \cite{wkleb} the
$FSB$ in strangeness decay constant was not taken into account, and this led to
underestimation of the strangeness excitation energies. Heavy
flavors $(c,b)$ have not been considered in these papers.

The addition of the term $L_6$ into starting lagrangian (\ref{L}) leads to modification
of the flavored moment of inertia, according to simple relation
 $\Theta_F = \Theta_F^{kin} + \Theta_F^{SK_4} + \Theta_F^{SK_6}$
.
But in order to take into account the symmetry breaking terms adequitly,
we have to express (in some order of $N_c^{-1}$) first set of coordinates
 (\ref{Omega})
in terms of collective coordinates $A(t)$ and $S(t)$ and substitute into $L_{rot}$.

The terms of the order of $N_c^{-1}$ in the hamiltonian, which  depend
 also
on the angular velocities of rotations in the isospin and the usual space
, are not crucial 
but important
 for the numerical estimates of the spectra of baryonic systems.
To calculate them one should first obtain the lagrangian of $BS$ including
all the terms upto $O(1/N_c)$. It can be written in
 a compact form, slightly different from
that given in \cite{wkleb},  as \cite{ksr}:

\bea
\label{L1}
L & \simeq &-M_{cl}+2\Theta_{F,B} \biggl[2\dot{D}^{\dagger}\dot{D} \biggl(1-
{d^2 \over 3} \biggr) - {4 \over 3}\bigl(D^\dagger \dot{D} \dot{D}^\dagger D-
(D^\dagger\dot{D})^2-(\dot{D}^\dagger D)^2\bigr) + (\vec{\omega}\vec{\beta})
\biggr] + \nonumber \\
 & +  & {\Theta_{I,B} \over2}
(\vec{\omega}-\vec{\beta})^2
 - \biggl[\Gamma_B\tilde{m}_D^2+(F_D^2-F_\pi^2)
\tilde{\Gamma}_B\biggr] D^\dagger D \biggl(1 - {d^2 \over 3}\biggr) + \nonumber\\
& + &i {N_cB \over 2}
\biggl(1- {d^2 \over 3}\biggr)(\dot{D}^\dagger D- D^\dagger \dot{D}) - {N_cB
\over 2} \vec{\omega} D^\dagger \vec{\tau} D,
\eea
where  $d^2=2D^\dagger D$ and
\be \label{beta} 
\vec{\beta}=-i(\dot{D}^\dagger \vec{\tau} D - D^\dagger
 \vec{\tau} \dot{D}). \ee
As we mentioned already, 
the role of the term $L_6$ reduces to the modification of the flavored inertia $\Theta_F$ in
(\ref{L1}). It is remarkable property of the starting lagrangian including $L_6$, that only 
quadratic terms in $\Omega_a$ enter (\ref{L1}). To get this expression we used the connection
between components $\Omega_a$ and $D,\,\dot{D},\,\omega_i$:
$\Omega_4^2+...+\Omega_7^2 = 8\dot{D}^{\dagger}\dot{D} \bigl(1-
d^2/3 \bigr) - 
16\bigl(D^\dagger \dot{D} \dot{D}^\dagger D-
(D^\dagger\dot{D})^2-(\dot{D}^\dagger D)^2\bigr)/3 + 4(\vec{\omega}\vec{\beta}) $,
and the component $\Omega_8$ which defines the $WZW$ term contribution, $\Omega_8 = \sqrt{3}\biggl[
i \bigl(1- d^2/3\bigr) (D^\dagger \dot{D} - \dot{D}^\dagger D ) + \vec{\omega} D^\dagger \vec{\tau} 
D
\biggr]$.

Taking into account the terms $\sim 1/N_c$ we find that the canonical
 variable $\Pi$ conjugate 
to $D$ is
\bea
\label{pi}
\Pi & = &\frac{\partial L}{\partial\dot{D}^{\dagger}}=
4\Theta_{F,B} \biggl[\dot{D}\biggl(1-{d^2 \over 3} \biggr)-
{2 \over 3} D^{\dagger}\dot{D} \, D+{4 \over 3}\dot{D}^{\dagger}D \, D\biggr]\nonumber \\
& + &i(\Theta_{I,B}-2\Theta_{F,B})\vec{\omega}\vec{\tau} D-
i\Theta_{I,B}\vec{\beta}\vec{\tau}D +i {N_cB \over 2}\biggl(1-
{d^2 \over 3} \biggr) \, D .
\eea

From (\ref{L1}) the body-fixed isospin operator is:
\be \vec{I}\sp{bf}=\partial L / \partial
\vec{\omega}=\Theta_{I,B}\vec{\omega}+(2\Theta_{F,B}-\Theta_{I,B})
\vec{\beta} - {N_cB \over 2} D^\dagger \vec{\tau} D,
\ee
which can be written also as:
\be
 \vec{I}^{bf}= \Theta_I \vec{\omega}+\bigl(1
-\frac{\Theta_I}{2\Theta_F}\bigr) \vec{I}_F -\frac{N_cB \Theta_I}{4\Theta_F}
D^{\dagger}\vec{\tau}D
\ee
with the operator 
\be
\label{if}
\hat{\vec{I}}_F = {i\over 2} \bigl(D^\dagger \vec{\tau}\Pi - \Pi^\dagger \vec{\tau} D\bigr) = 
(b^{\dagger}\vec{\tau}b -
a^T\vec{\tau}a^{\dagger T})/2
\ee.
Using connection between $\Pi,\;\dot{D}$ and $D$ given by (\ref{pi}) in leading order
we obtain for the quantity $\beta$ in (\ref{beta})
\be
\label{bbeta}
\vec{\beta} \simeq {1\over 2\Theta_F}\biggl(\vec{I}_F + {N_cB\over 2} D^\dagger \vec{\tau} D\biggr)
\ee
For the states with definite flavor quantum number we should make substitution 
$D^\dagger \vec{\tau} D \to -2\vec{I}_F/(N_cB\mu_F)$ (for flavor) or
$D^\dagger \vec{\tau} D \to 2\vec{I}_F/(N_cB\mu_F)$ for antiflavor, and
we can write for matrix elements of states with definite flavor:
\be
\label{ibff}
\vec{I}^{bf}= \Theta_{I,B}\vec{\omega} + c_{F,B} \vec{I}_F,
\ee
with
\be
\label{cff}
c_{F,B}=1 - \frac{\Theta_{I,B}}{2\Theta_{F,B}\mu_{F,B}}(\mu_{F,B}-1).
\ee
We used also that within our approximation
\be
\Theta_{I,B}\vec{\beta} \simeq (1-c_{F,B})\vec{I}_F.
\ee
Relation, similar to (\ref{ibff}) holds also for antiflavor with
\be
  c_{\bar{F},B}=1 -\frac{\Theta_{I,B}}{2\Theta_{F,B}\mu_{F,B}}(\mu_{F,B}+1),
\ee
so, it differs from (\ref{cff}) by change $\mu \to -\mu$.
Using the identities :
\be -i \vec{\beta} \vec{\tau}D=2 D^\dagger D\dot{D}-(\dot{D}^\dagger
D+D^\dagger \dot{D}) D
\ee
and
\be \vec{\beta}^2 = 4 D^\dagger D \dot{D}^\dagger\dot{D}-(\dot{D}^\dagger D
+D^\dagger \dot{D})^2
\ee
we find  that
 the $\sim 1/N_c$ zero mode quantum corrections to the energies of skyrmions
can be estimated \cite{wkleb} as:
\be
\label{1/N_c}
\Delta E_{1/N_c} = {1 \over 2\Theta_{I,B}}\bigl[c_{F,B} I_r(I_r+1)+
(1-c_{F,B})I(I+1) + (\bar{c}_{F,B}-c_{F,B})I_F(I_F+1) \bigr],
\ee
where $I=I^{bf}$ is the value of the isospin of the baryon or $BS$.
$I_r$ is the quantity analogous to the
``right" isospin $I_r$, in the collective coordinate  approach \cite{g},
and $\vec{I_r}=\vec{I}^{bf}-\vec{I_F}$. The hyperfine structure constants
$c_{F,B}$ are given in (\ref{cff}), and $\bar{c}_{F,B}$ are defined by relations:
\be 
\label{anticf}
1 - \bar{c}_{F,B}=\frac{\Theta_{I,B}}{\Theta_{F,B}(\mu_{F,B})^2}(\mu_{F,B}-1), \quad
1 - \bar{c}_{\bar{F},B}=-\frac{\Theta_{I,B}}{\Theta_{F,B}(\mu_{F,B})^2}(\mu_{F,B}+1).
\ee
For nucleons $I=I_r=1/2$, $I_F=0$ and $\Delta E_{1/N_c}(N) =3/(8\Theta_{I,1})$,
for $\Delta$ - isobar $I=I_r=3/2, \; I_F=0$, and $\Delta E_{1/N_c}(\Delta) =15/(8\Theta_{I,1})$,
as in $SU(2)$ quantization scheme. The $\Delta-N$ mass splitting is described satisfactorily,
according to values of $\Theta_I$ presented in {\bf Table 1}.

\begin{center}
\tenrm
\begin{tabular}{|l|l|l|l|l|l|l|l|l|l|l|l|l|}
\hline
$B$ & $\omega_s^{SK_4}$ & $\omega_c^{SK_4}$ & $\omega_b^{SK_4}$ & $\omega_s^{SK_6}$ &
$\omega_c^{SK_6}$ & $\omega_b^{SK_6}$ & $\omega_s^{SK_4^*}$ & $\omega_c^{SK_4^*}$ &
$\omega_b^{SK_4^*}$ & $\omega_s^{SK_6^*}$ & $\omega_c^{SK_6^*}$ & $\omega_b^{SK_6^*}$ \\ \hline
$1 $ & $0.307$ & $1.54$ & $4.80$ & $0.336$ & $1.61$ & $4.93$ & $0.345$ & $1.55$ & $4.77$ & $0.375$ & $1.62$ & $4.89$ \\ \hline
$2 $ & $0.298$ & $1.52$ & $4.77$ & $0.346$ & $1.64$ & $4.98$ & $0.339$ & $1.54$ & $4.75$ & $0.386$ & $1.66$ & $4.95$ \\ \hline
$3 $ & $0.293$ & $1.51$ & $4.76$ & $0.342$ & $1.64$ & $4.98$ & $0.336$ & $1.54$ & $4.74$ & $0.385$ & $1.66$ & $4.95$ \\ \hline
$4 $ & $0.285$ & $1.50$ & $4.74$ & $0.328$ & $1.62$ & $4.95$ & $0.330$ & $1.52$ & $4.72$ & $0.377$ & $1.64$ & $4.93$ \\ \hline
$5 $ & $0.290$ & $1.51$ & $4.75$ & $0.334$ & $1.63$ & $4.96$ & $0.334$ & $1.53$ & $4.74$ & $0.380$ & $1.65$ & $4.94$ \\ \hline
$6 $ & $0.290$ & $1.51$ & $4.76$ & $0.332$ & $1.63$ & $4.96$ & $0.334$ & $1.54$ & $4.74$ & $0.379$ & $1.65$ & $4.94$ \\ \hline
$7 $ & $0.285$ & $1.50$ & $4.74$ & $0.324$ & $1.62$ & $4.95$ & $0.331$ & $1.53$ & $4.73$ & $0.374$ & $1.64$ & $4.93$ \\ \hline
$8 $ & $0.290$ & $1.51$ & $4.76$ & $0.329$ & $1.63$ & $4.96$ & $0.335$ & $1.54$ & $4.75$ & $0.377$ & $1.65$ & $4.94$ \\ \hline
$9 $ & $0.292$ & $1.52$ & $4.77$ & $0.331$ & $1.63$ & $4.97$ & $0.336$ & $1.54$ & $4.76$ & $0.378$ & $1.65$ & $4.94$ \\ \hline
$10$ & $0.293$ & $1.52$ & $4.78$ & $0.331$ & $1.63$ & $4.97$ & $0.337$ & $1.55$ & $4.76$ & $0.378$ & $1.65$ & $4.94$ \\ \hline
$11$ & $0.295$ & $1.53$ & $4.79$ & $0.332$ & $1.63$ & $4.97$ & $0.338$ & $1.55$ & $4.77$ & $0.378$ & $1.65$ & $4.95$ \\ \hline
$12$ & $0.295$ & $1.53$ & $4.79$ & $0.331$ & $1.63$ & $4.97$ & $0.338$ & $1.55$ & $4.77$ & $0.378$ & $1.65$ & $4.95$ \\ \hline
$13$ & $0.296$ & $1.53$ & $4.79$ & $0.332$ & $1.63$ & $4.98$ & $0.339$ & $1.55$ & $4.77$ & $0.378$ & $1.65$ & $4.95$ \\ \hline
$14$ & $0.300$ & $1.54$ & $4.80$ & $0.335$ & $1.64$ & $4.98$ & $0.342$ & $1.56$ & $4.79$ & $0.379$ & $1.65$ & $4.95$ \\ \hline
$15$ & $0.301$ & $1.54$ & $4.81$ & $0.336$ & $1.64$ & $4.99$ & $0.343$ & $1.56$ & $4.79$ & $0.380$ & $1.66$ & $4.95$ \\ \hline
$16$ & $0.302$ & $1.54$ & $4.81$ & $0.336$ & $1.64$ & $4.99$ & $0.343$ & $1.56$ & $4.79$ & $0.380$ & $1.66$ & $4.96$ \\ \hline
$17$ & $0.302$ & $1.54$ & $4.81$ & $0.335$ & $1.64$ & $4.99$ & $0.343$ & $1.56$ & $4.79$ & $0.379$ & $1.66$ & $4.95$ \\ \hline
$20$ & $0.308$ & $1.56$ & $4.84$ & $0.340$ & $1.65$ & $5.00$ & $0.347$ & $1.58$ & $4.81$ & $0.382$ & $1.66$ & $4.96$ \\ \hline
$24$ & $0.312$ & $1.57$ & $4.85$ & $0.343$ & $1.66$ & $5.01$ & $0.351$ & $1.58$ & $4.83$ & $0.384$ & $1.66$ & $4.97$ \\ \hline
$28$ & $0.316$ & $1.58$ & $4.87$ & $0.347$ & $1.66$ & $5.02$ & $0.354$ & $1.59$ & $4.85$ & $0.385$ & $1.67$ & $4.98$ \\ \hline
$32$ & $0.319$ & $1.59$ & $4.88$ & $0.349$ & $1.67$ & $5.02$ & $0.356$ & $1.60$ & $4.86$ & $0.386$ & $1.67$ & $4.98$ \\ \hline
\end{tabular}
\end{center}

{\bf Table 3.}
 {\small Flavor excitation energies for strangeness, charm and beauty, in $Gev$. $e=4.12$ for the $SK4$ variant
and $e'=4.11$ for the $SK6$ model. For rescaled variants (numbers marked with $^*$) $e=3.00$ and
$e'=2.84$ for $SK4$ and $SK6$ variants, correspondingly.
The ratio $F_D/F_\pi =1.5$ for charm, and $F_B/F_\pi = 2$ for beauty.}\\

As can be seen from {\bf Table 3}, for the $SK4$ variant there is some decrease of flavor excitation
energies when $B$-number increases from $1$ to $7$, but further these energies increase again,
and for $B \geq 20$ they exceed the $B=1$ value. The latter property can be connected, however,
with specifics of the rational map approximation (the quantity $\Gamma$ increases faster than 
the flavored inertia $\Theta_F$, see (\ref{omdif})) which becomes less realistic for larger values of
$B$. Such behaviour of $\omega$-s is important for conclusions about possible existence of
hypernuclei \cite{vk3}. The {\bf Table 3} is presented here for comparison with antiflavor excitation 
energies presented in {\bf Table 4}.
 Generally, the rigid oscillator version of the bound state
model we are using here, overestimates the flavor excitation energies. However, phenomenological
consequences derived in \cite{vk1,vk3} for the binding energies of strange $S=-1$ hypernuclei are 
based mainly on the differences of these energies. The qualitative and in some cases quantitative
agreement takes place between data for binding energies of ground states of hypernuclei with atomic
numbers between $5$ and $20$ and results of calculations within $SK4$ variant of the chiral soliton
model taking into account collective motion of solitons in $SU(3)$ configuration space \cite{vk3}.

Another peculiarity of interest is that for rescaled variant of the model the charm and 
beauty excitation energies are very close to those of original variant (scaling property), but 
differ more substantially for strangeness - greater by $\sim 30\,-\,40\,Mev$. Such somewhat 
unexpected behaviour is connected with the fact that flavor excitation energies appear as a result of
subtraction of two quantities, which behave differently when rescaling is made, see (\ref{om}).

\begin{center}
\tenrm
\begin{tabular}{|l|l|l|l|l|l|l|l|l|l|l|l|l|}
\hline
$B$ & $\bar{\omega}_s^{SK_4}$ & $\bar{\omega}_c^{SK_4}$ & $\bar{\omega}_b^{SK_4}$ &
$\bar{\omega}_s^{SK_6}$ & $\bar{\omega}_c^{SK_6}$ & $\bar{\omega}_b^{SK_6}$ &
$\bar{\omega}_s^{SK_4^*}$ & $\bar{\omega}_c^{SK_4^*}$ & $\bar{\omega}_b^{SK_4^*}$ &
$\bar{\omega}_s^{SK_6^*}$ & $\bar{\omega}_c^{SK_6^*}$ & $\bar{\omega}_b^{SK_6^*}$ \\ \hline
$1 $ & $0.591$ & $1.75$ & $4.94$ & $0.584$ & $1.79$ & $5.04$ & $0.472$ & $1.65$ & $4.83$ & $0.468$ & $1.69$ & $4.93$ \\ \hline
$2 $ & $0.571$ & $1.72$ & $4.90$ & $0.571$ & $1.80$ & $5.08$ & $0.459$ & $1.63$ & $4.81$ & $0.470$ & $1.72$ & $4.99$ \\ \hline
$3 $ & $0.564$ & $1.71$ & $4.89$ & $0.569$ & $1.80$ & $5.07$ & $0.455$ & $1.63$ & $4.80$ & $0.468$ & $1.72$ & $4.99$ \\ \hline
$4 $ & $0.567$ & $1.71$ & $4.87$ & $0.580$ & $1.80$ & $5.06$ & $0.454$ & $1.62$ & $4.78$ & $0.468$ & $1.71$ & $4.97$ \\ \hline
$5 $ & $0.558$ & $1.71$ & $4.88$ & $0.571$ & $1.80$ & $5.07$ & $0.452$ & $1.62$ & $4.80$ & $0.466$ & $1.71$ & $4.98$ \\ \hline
$6 $ & $0.555$ & $1.71$ & $4.88$ & $0.571$ & $1.80$ & $5.07$ & $0.451$ & $1.62$ & $4.80$ & $0.465$ & $1.71$ & $4.98$ \\ \hline
$7 $ & $0.559$ & $1.71$ & $4.88$ & $0.578$ & $1.80$ & $5.06$ & $0.451$ & $1.62$ & $4.79$ & $0.466$ & $1.71$ & $4.97$ \\ \hline
$8 $ & $0.553$ & $1.71$ & $4.89$ & $0.571$ & $1.80$ & $5.07$ & $0.450$ & $1.63$ & $4.80$ & $0.465$ & $1.71$ & $4.98$ \\ \hline
$9 $ & $0.550$ & $1.71$ & $4.90$ & $0.569$ & $1.80$ & $5.07$ & $0.450$ & $1.63$ & $4.81$ & $0.465$ & $1.71$ & $4.98$ \\ \hline
$10$ & $0.549$ & $1.71$ & $4.90$ & $0.569$ & $1.80$ & $5.07$ & $0.450$ & $1.63$ & $4.82$ & $0.465$ & $1.71$ & $4.98$ \\ \hline
$11$ & $0.547$ & $1.71$ & $4.90$ & $0.567$ & $1.80$ & $5.08$ & $0.450$ & $1.63$ & $4.82$ & $0.464$ & $1.71$ & $4.98$ \\ \hline
$12$ & $0.547$ & $1.72$ & $4.91$ & $0.568$ & $1.80$ & $5.08$ & $0.450$ & $1.63$ & $4.82$ & $0.464$ & $1.71$ & $4.98$ \\ \hline
$13$ & $0.546$ & $1.72$ & $4.91$ & $0.567$ & $1.80$ & $5.08$ & $0.450$ & $1.64$ & $4.83$ & $0.464$ & $1.71$ & $4.99$ \\ \hline
$14$ & $0.543$ & $1.72$ & $4.92$ & $0.564$ & $1.80$ & $5.08$ & $0.450$ & $1.64$ & $4.84$ & $0.464$ & $1.72$ & $4.99$ \\ \hline
$15$ & $0.542$ & $1.72$ & $4.92$ & $0.563$ & $1.80$ & $5.08$ & $0.450$ & $1.64$ & $4.84$ & $0.464$ & $1.72$ & $4.99$ \\ \hline
$16$ & $0.541$ & $1.72$ & $4.93$ & $0.562$ & $1.80$ & $5.08$ & $0.450$ & $1.64$ & $4.85$ & $0.464$ & $1.72$ & $4.99$ \\ \hline
$17$ & $0.542$ & $1.72$ & $4.93$ & $0.564$ & $1.80$ & $5.09$ & $0.450$ & $1.64$ & $4.85$ & $0.464$ & $1.72$ & $4.99$ \\ \hline
$18$ & $0.540$ & $1.72$ & $4.93$ & $0.561$ & $1.81$ & $5.09$ & $0.451$ & $1.65$ & $4.85$ & $0.464$ & $1.72$ & $5.00$ \\ \hline
$19$ & $0.539$ & $1.73$ & $4.94$ & $0.559$ & $1.81$ & $5.09$ & $0.451$ & $1.65$ & $4.86$ & $0.464$ & $1.72$ & $5.00$ \\ \hline
$20$ & $0.538$ & $1.73$ & $4.94$ & $0.558$ & $1.81$ & $5.09$ & $0.451$ & $1.65$ & $4.86$ & $0.464$ & $1.72$ & $5.00$ \\ \hline
$24$ & $0.536$ & $1.73$ & $4.96$ & $0.555$ & $1.81$ & $5.10$ & $0.452$ & $1.66$ & $4.88$ & $0.463$ & $1.72$ & $5.00$ \\ \hline
$28$ & $0.533$ & $1.74$ & $4.97$ & $0.552$ & $1.81$ & $5.10$ & $0.453$ & $1.67$ & $4.89$ & $0.463$ & $1.72$ & $5.01$ \\ \hline
$32$ & $0.532$ & $1.74$ & $4.98$ & $0.550$ & $1.81$ & $5.11$ & $0.453$ & $1.67$ & $4.90$ & $0.463$ & $1.73$ & $5.01$ \\ \hline
\end{tabular}
\end{center}

{\bf Table 4.}
 {\small Antiflavor excitation energies for strangeness, charm and beauty, as in {\bf Table 3}.
In original variants of the model $e=4.12$ for the $SK4$ variant and $e'=4.11$ for the $SK6$
variant. The numbers with $^*$ are for the rescaled variants of the model, $e=3.0$ for the $SK4$ 
variant and $e'=2.84$ for the $SK6$ variant.
 The ratio $F_D/F_\pi =1.5$ for charm and $F_B/F_\pi = 2$ for beauty.}\\

Similar to flavor energies, there is remarkable universality of antiflavor excitation energies for 
different baryon numbers, especially
 for anti-charm and anti-beauty: variations do not exceed few $\%$.
It follows from {\bf Table 4} that there is some decrease of antiflavor excitation energies when
$B$ increases from $1$, this effect is striking for the $SK4$ variant and especially for
strangeness. Within the $SK6$ variant the $B=1$ energies for anti-charm 
and anti-beauty  are slightly 
smaller than for $B\geq 2$.

For the case of strangeness, the $\bar{\omega}_s$ decreases with increasing B-number in most cases,
as it can be seen from {\bf Table 4} (except the rescaled $SK6$ variant where $B=1$ energy is slightly
smaller than $B=2$ one), but it is always greater than kaon mass, therefore, the state with positive 
strangeness can decay strongly into kaon and some final nucleus, or nuclear fragments.

The heavy antiflavors excitation energies also reveal notable scale independence,
i.e. the values obtained with constant $e=4.12$ and $3.00$ ($SK4$ variant) shown in {\bf Tables 3,4}, 
are close to each other within several percents, as well as values for $e'=4.11$ and $2.84$ for 
the $SK6$ variant. It was really  expected from general
 arguments for large values of $FSB$ meson 
mass \cite{kz}.
 The change of numerical values of these energies is, however, important for 
conclusions concerning the binding energies of nuclear states with antiflavors.

All excitation energies of antiflavors are smaller for rescaled variants, i.e. when constants
$e$ or $e'$ are decreased by about $30\%$. It seems to be natural since dimensions of multiskyrmions 
which scale like $1/(F_\pi e)$ increase due to this change, and all energies become "softer".
Such behaviour is due to the fact that antiflavor energies are the sum of two terms
(see above (\ref{om})) which behave (roughly!) in similar manner when rescaling is made. Remarkably, 
the 
decrease of energies due to rescaling is of the order of $\sim 100\,Mev$ in all cases 
(e.g., for anti-strangeness and $B=1$ it is $119\,Mev$ ($SK4$ variant) and $116\,Mev$ in $SK6$ variant),
and slightly smaller for $\bar{c}$ (decrease due to rescaling about $\sim 100\,Mev$) and $\bar{b}$ 
(decrease by $110\,Mev$).
\section{The binding energies of $\Theta^+$-hypernuclei and anti-charmed (anti-beautiful) hypernuclei}

In view of the large enough values of anti-strangeness excitation energies one 
cannot speak about positive strangeness
 hypernuclei, which decay weakly, similar to ordinary $S=-1$ 
hypernuclei. However, one can speak about $\Theta$-hypernuclei, where $\Theta$-hyperon is bound by
several nucleons. One of puzzling properties of pentaquarks is their small width, $\Gamma_\Theta <
\sim 10 \,Mev$ according to experiments where $\Theta^+$ has been observed \cite{japan,step}, and
probably even smaller, according to analyses of kaon-nucleon interaction data 
\cite{aps}. Possible explanations, from some numerical cancellation \cite{dpp} and cancellation in large 
$N_c$ expansion \cite{mic3} to qualitative one in terms of the quark model wave function \cite{kl,jw}, 
and calculation using operator product expansion \cite{io} have been proposed \footnote{In most of
variants of explanation it is difficult to expect the width of $\Theta$-hyperon of the order of
$\sim 1\,Mev$, as obtained in \cite{aps}. Therefore, the checking of data analyzed in \cite{aps} seems
to be of first priority.}.
The width of nuclear
 bound states of $\Theta$ should be of same order of magnitude as the width of 
$\Theta^+$ itself, or smaller: besides the smaller energy release some suppression due to the Pauli blocking 
for the final
 nucleon from $\Theta$ decay can take place for heavier nuclei.

For anti-charm and anti-beauty the excitation energies are smaller than masses of $D$- or $B$-
meson, and it makes sense to consider the possibility of existence of anti-charmed or anti-beauty
hypernuclei which have the time of life characteristic for weak interactions.

In the bound state soliton model, and in its rigid oscillator version as
well, the states predicted do not correspond apriory to
 the definite $SU(3)$ or $SU(4)$ 
representations.
They can be ascribed to definite $SU(3)$ irreps, as it was shown in \cite{wkleb,kz}. Due to
configuration mixing caused by the flavor symmetry breaking, each state with
definite value of flavor, $s,\;c$ or $b$, is some mixture of the components of several $SU(3)$
irreps with given value of $F$ and isospin $I$ which is strictly conserved in our approach
(until explicitly isospin violating terms are included into lagrangian). In the case of strangeness,
as calculations show (see e.g. \cite{kss}), this mixture is dominated usually by the lowest $SU(3)$ 
irrep, and admixtures
do not exceed several percents, usually. Situation is changed for charm or beauty quantum
numbers, where admixtures can have weight comparable with the weight of the lowest configuration.
However, we consider here the simplest possibility of one lowest irrep, for rough estimates.

Let $(p,q)$ characterize the $SU(3)$ irrep to which $BS$ belongs, then quantization condition 
$(p+2q)=N_cB$ \cite{g}, for arbitrary $N_c$,  changes to
$(p+2q)=N_cB+3m$, where $m$ is related to the number of additional  quark-antiquark pairs
$n_{q\bar{q}}$ present in the quantized states, $n_{q\bar{q}} \geq m$  \cite{vk,jenkm}.
The non-exotic states with $m=0$, or minimal states, have $p+2q = 3B$, ($N_c=3$ further), and 
states with lowest "right" isospin $I_r=p/2$ have $(p,q)=(0,3B/2)$ for even $B$, and 
$(p,q)=(1,(3B-1)/2)$ for odd $B$ \cite{vk,kss}. E.g., the state with $B=1$, $|F|=1$, $I=0$ and
$n_{q\bar{q}}=0$ should belong to the octet of $(u,d,s)$, or $(u,d,c)$,
 $SU(3)$ group, if $N_c=3$; see also \cite{wkleb}. For the first exotic states the lowest possible $SU(3)$ irreps $(p,q)$ for each value of the baryon 
number $B$ are defined by
 relations:
$p+2q=3(B+1)$; $p=1, \; q=(3B+2)/2$ for even $B$, and $p=0, \; q=(3B+3)/2$ for odd $B$.
E.g., for $B=2, \, 4, \, 6$ and $8$ we have $\overline{35}$, $\overline{80}$,  $\overline{143}$ and
$\overline{224}$-plets, 
and for $B=3,\, 5$ and $7$ - multiplets $\overline{28}, \, \overline{55}$ 
and $\overline{91}$, etc.

Since we are interested in the lowest energy states, we discuss here the
baryonic systems with the lowest allowed angular momentum, {\it ie} $J=0$,
for $B=4, \; 6$ etc. and $J=1/2$ for odd values of the $B$-number. There are some 
deviations from this simple law for the ground states of real nuclei, but anyway,
the correction to the energy of quantized states due to collective rotation of 
solitons is small and decreases with increasing $B$ since the corresponding
moment of inertia increases proportionally to $\sim B^2$ \cite{vk1,ash}.
Moreover, the $J$-dependent correction to the energy may cancel in the
differences of energies of flavored and flavorless states, so, we neglect these contributions
for our rough estimates. 

For the non-exotic states we considered previously \cite{vk3} the energy difference between the 
state with flavor $F$
 belonging to the $(p,q)$ irrep, and the ground state with $F=0$ and the same 
angular momentum
 and $(p,q)$. Situation is different for exotic states, since exotic and non-exotic 
states have
 different values of $(p,q)$. The difference between $\bar{\omega}$ and $\omega$,
$\bar{\omega} - \omega = N_c/(4\Theta_F)$ takes into account this distinction in the values of
$(p,q)$, as it is shown explicitly in Appendix.

For the case of $B=1,\;3,\;5,...$ etc. we have for the ground state $I=I_r=1/2,\;I_F=0$, therefore,
the correction  $\Delta E_{1/N_c} = I(I+1)/(2\Theta_{I,B}) =3/(8\Theta_{I,B})$. For exotic
antiflavored state we have $I=0$, $I_r=I_F=1/2$, and correction equals to
$\Delta E_{1/N_c} = 3\bar{c}_{F,B}/8\Theta_{I,B}$.
For the difference of energies between exotic and non-exotic ground states we obtain:
\bea
\label{dele}
\Delta E_{B,F}& = &\bar{\omega}_{F,B} + \frac{3(\bar{c}_{\bar{F},B}-1)}{8\Theta_{F,B}} 
= \bar{\omega}_{F,B}+  \frac{3(\mu_{F,B}+1)}{8\mu_{F,B}^2 \Theta_{F,B}},
\eea
Note that the moment of inertia $\Theta_I$ does not enter the difference of energies (\ref{dele}).

For the $B=1$ case, the difference of masses
 of $\Theta_F$ and the nucleon is
\bea \Delta M_{\Theta_FN} =  \overline{\omega}_{F,1}- \frac{3(1-\bar{c}_{\bar{F},1})}
{8\Theta_{I,1}}= \overline{\omega}_{F,1}+\frac{3(\mu_{F,1}+1)}{8\mu_{F,1}^2\Theta_{F,1}},
\eea
The difference of masses of $\Theta$ and $\Lambda$ hyperons also is of interest and can be presented
in simple form:
\be
\label{dtl}
 \Delta M_{\Theta_F\Lambda_F} =  \overline{\omega}_{F,1}- \omega_{F,1}
+ \frac{3(\bar{c}_{\bar{F},1}-\bar{c}_{F,1})}
{8\Theta_{I,1}}= 
 \frac{3(\mu_{F,1}+1)}{4\mu_{F,1}\Theta_{F,1}}.
\ee

For the case of even $B = 4,\;6,...$ etc. the ground state has $I=I_r=I_F=0$ (like for nucleus
$^4 He$), and $\Delta E_{1/N_c} =0$. For the first exotic states $I=I_F=1/2$, and we have a choice
for $I_r$, $I_r=0$ or $1$. If $c_{\bar{F},B} = 1 - \Theta_{I,B}(\mu_{F,B}+1)/(2\Theta_{F,B}\mu_{F,B})
>0$, $I_r=0$, if $c_{\bar{F},B} <0$, we should take  $I_r=1$.
In the first case, the correction to the energy of the state
$\Delta E_{1/N_c} = 3(1+\bar{c}_{\bar{F},B}-2c_{\bar{F},B})/8\Theta_{I,B} = 3 (\mu_{F,B} +1)^2/
(8\Theta_{F,B}\mu_{F,B}^2)$.

\begin{center}
\begin{tabular}{|l|l|l|l|l|l|l|l|l|}
\hline
$B$ & $ \Delta\epsilon^{SK_4}$ & $\epsilon^{SK_4}$ &$\Delta\epsilon^{SK_6}$ &$\epsilon^{SK_6}$ & 
$\Delta\epsilon^{SK_4^*}$ & $\epsilon^{SK_4^*}$& $\Delta\epsilon^{SK_6^*}$&$ \epsilon^{SK_6^*}
$\\
\hline
2  &  47   &   47  &   75  &    75  &   25   &    25   &   17   &    17
\\
3  &  67   &   76  &   45  &    53  &   26   &    34   &   4    &    12
\\
4  &  20   &   49  &   -4  &    24  &   9    &    38   &   -8   &    20
\\
5  &  81   &   108 &   47  &    74  &   30   &    57   &   6    &    33
\\
6  &  56   &   88  &   24  &    56  &   20   &    52   &   -1   &    31
\\
7  &  83   &   121 &   41  &    80  &   32   &    70   &   7    &    45
\\
8  &  69   &   126 &   31  &    87  &   24   &    81   &   2    &    58
\\
9  &  94   &   152 &   53  &    110 &   33   &    90   &   8    &    66
\\
10 &  79   &   144 &   39  &    103 &   27   &    92   &   4    &    68
\\
11 &  99   &   173 &   56  &    130 &   33   &    108  &   9    &    84
\\
12 &  86   &   178 &   43  &    135 &   28   &    120  &   5    &    97
\\
13 &  101  &   196 &   56  &    152 &   33   &    129  &   9    &    104
\\
14 &  93   &   197 &   50  &    154 &   29   &    133  &   6    &    111
\\
15 &  105  &   219 &   61  &    175 &   33   &    147  &   9    &    123
\\
16 &  96   &   224 &   53  &    181 &   29   &    157  &   7    &    134
\\
17 &  105  &   235 &   61  &    191 &   33   &    163  &   9    &    139
\\
18 &  100  &   237 &   56  &    194 &   29   &    167  &   7    &    144
\\
19 &  109  &   255 &   65  &    211 &   33   &    178  &   10   &    156
\\
20 &  103  &   263 &   60  &    220 &   29   &    190  &   8    &    168
\\
21 &  111  &   276 &   67  &    232 &   32   &    197  &   10   &    175
\\
22 &  105  &   279 &   62  &    236 &   29   &    203  &   8    &    182
\\
23 &  113  &   297 &   69  &    253 &   32   &    216  &   10   &    194
\\
24 &  107  &   305 &   64  &    263 &   29   &    228  &   8    &    206
\\
25 &  113  &   316 &   70  &    273 &   31   &    235  &   10   &    213
\\
26 &  109  &   321 &   66  &    278 &   29   &    241  &   8    &    220
\\
27 &  115  &   337 &   72  &    294 &   31   &    253  &   10   &    232
\\
28 &  111  &   347 &   69  &    305 &   29   &    265  &   9    &    245
\\
29 &  116  &   358 &   73  &    315 &   31   &    273  &   10   &    252
\\
30 &  112  &   363 &   70  &    321 &   29   &    279  &   9    &    259
\\
31 &  117  &   376 &   75  &    335 &   30   &    290  &   10   &    270
\\
32 &  113  &   385 &   71  &    343 &   29   &    300  &   9    &    281
\\
\hline

\end{tabular}
\end{center}
{\bf Table 5.} {\small The binding energy differences and total binding energies of positive 
strangeness $\Theta^+$-hypernuclei (in $Mev$) for the $SK4$ and $SK6$ variants of the model in 
rational map approximation.}\\

The binding energy differences $\Delta \epsilon_{\bar{s},\bar{c},\bar{b}}$ are the changes
of binding energies of lowest $BS$ with flavor $\bar{s},\,\bar{c}$ or $\bar{b}$ and isospin
$I=0$ (for odd $B$) and $I=1/2$ (for even $B$) in comparison with usual $u,d$ nuclei
(when one nucleon is replaced by $\Theta$-hyperon - in other words). The classical masses of
skyrmions are cancelled in such differences:
\be
\Delta \epsilon_{B,F} = \Delta E_{gr.st.}(B) - \Delta E (B,F) + \Delta M_{\Theta_FN}.
\ee
It follows from (\ref{dele})  that this change of the binding energy of the system is, for odd
$B$-number
\be
\label{delodd}
\Delta \epsilon_{B,F} = \bar{\omega}_{F,1}-\bar{\omega}_{F,B} +
\frac{3(\mu_{F,1}+1)}{8\mu_{F,1}^2\Theta_{F,1}} - \frac{3(\mu_{F,B}+1)}{8\mu_{F,B}^2\Theta_{F,B}}.
\ee
Evidently, in the limit of very heavy flavor, $\mu_F \to \infty$,
\be
\Delta \epsilon_{B,F} \to  \bar{\omega}_{F,1}-\bar{\omega}_{F,B}.
\ee
For $B$-numbers $4,\,6,...$ etc. we obtain

\be
\label{deleven}
\Delta \epsilon_{B,F} = \bar{\omega}_{F,1}-\bar{\omega}_{F,B} +
\frac{3(\mu_{F,1}+1)}{8\mu_{F,1}^2\Theta_{F,1}}
-  \frac{3(\mu_{F,B}+1)^2}{8\mu_{F,B}^2\Theta_{F,B}}.
\ee
In the limit of very heavy flavor we have from (\ref{deleven})
\be
\Delta \epsilon_{B,F} = \bar{\omega}_{F,1}-\bar{\omega}_{F,B} -
  \frac{3}{8\Theta_{F,B}},
\ee
so, in comparison with the case of odd $B$-numbers there is additional contribution
decreasing with increase of the $B$-number (because inertia increase with $B$) from $\sim 25\;Mev$
for $B=3$.
 Numerical estimates for the total binding
 energies of anti-flavored states, presented in 
{\bf Tables 5, 6, 7}, are obtained by adding the values of (\ref{delodd}, \ref{deleven}) 
to the binding energies of ordinary nuclei we take from existing data.

The case of $B=2$ is a special one because the equality $I=J=0$ is forbidden for two-nucleon system
in $S$-wave due to the Pauli principle. In {\bf Tables 5-7} we present the binding energy estimates
relative to $NN$-scattering state (quasideuteron) with isospin $I=1$. To get them, we added
the values $1/\Theta_{I,B=2}$ to the number given by (\ref{deleven}).

One should keep in mind that for the $SK4$ model the value of $\Theta^+$ mass equals to $1588\,Mev$,
i.e. by $\sim 150\,Mev$ above kaon-nucleon threshold. By this reason the states with biggest
binding energy shown in {\bf Table 5} are unstable relative to strong interactions.
For the $SK6$ variant $M_\Theta=1566\,Mev$, and the binding energies are considerably smaller, by
$\sim 40-50\,Mev$ in some cases - this is the main feature of the $SK6$ variant. For the rescaled 
variants the difference between both variants is reduced considerably, but in this case the binding
energies are underestimated.

From phenomenological point of view we should describe the $B=1$ states with original variants
of models, i.e. $e=4.12,\; e'=4.11$ and states with $\sim 10 < B=A < \sim 30$  -
using rescaled variants, as it is suggested by results of \cite{ksm}. This procedure gives
most optimistic values of $\Delta \epsilon_{S=+1}$, about $145\,Mev$ for the $SK4$ variant and
$\sim 140\,Mev$ for the $SK6$ variant. However, uncertainty of this prediction is few tens of $Mev$,
at least. 

\begin{center}
\begin{tabular}{|l|l|l|l|l||l|l|l|l|l|l|l|l|l|}
\hline
$B$
& $\Delta \epsilon_{\bar{c}}^{SK4}$& $\epsilon_{\bar{c}}$& $\Delta \epsilon_{\bar{b}}^{SK4}$&
$\epsilon_{\bar{b}}$ & $\Delta \epsilon_{\bar{c}}^{SK6}$&
$\epsilon_{\bar{c}}$& $\Delta \epsilon_{\bar{b}}^{SK6}$& $\epsilon_{\bar{b}}$ \\
\hline
$2 $ & $61$  & $61 $ & $91 $ & $91 $ & $56 $ & $56 $ & $ 44$ & $ 44$ \\
$3 $ & $38 $ & $46 $ & $49 $ & $57 $ & $-8 $ & $0  $ & $-36$ & $-28$ \\
$4 $ & $15 $ & $44 $ & $48 $ & $76 $ & $-29$ & $-1 $ & $-36$ & $-7 $ \\
$5 $ & $44 $ & $71 $ & $55 $ & $82 $ & $-5 $ & $22 $ & $-30$ & $-3 $ \\
$6 $ & $27 $ & $59 $ & $43 $ & $75 $ & $-20$ & $12 $ & $-39$ & $-7 $ \\
$7 $ & $47 $ & $85 $ & $62 $ & $101$ & $-5 $ & $34 $ & $-23$ & $16 $ \\
$8 $ & $31 $ & $87 $ & $41 $ & $98 $ & $-17$ & $40 $ & $-37$ & $19 $ \\
$9 $ & $42 $ & $100$ & $43 $ & $100$ & $-6 $ & $51 $ & $-33$ & $24 $ \\
$10$ & $31 $ & $96 $ & $33 $ & $98 $ & $-15$ & $50 $ & $-40$ & $25 $ \\
$11$ & $40 $ & $114$ & $34 $ & $108$ & $-7 $ & $68 $ & $-37$ & $37 $ \\
$12$ & $31 $ & $123$ & $27 $ & $119$ & $-15$ & $78 $ & $-42$ & $50 $ \\
$16$ & $27 $ & $154$ & $8  $ & $136$ & $-15$ & $113$ & $-50$ & $78 $ \\
$17$ & $32 $ & $162$ & $11 $ & $141$ & $-10$ & $120$ & $-47$ & $83 $ \\
$20$ & $22 $ & $183$ & $-7 $ & $154$ & $-15$ & $145$ & $-57$ & $104$ \\
$24$ & $19 $ & $217$ & $-19$ & $179$ & $-16$ & $182$ & $-62$ & $136$ \\
$28$ & $15 $ & $251$ & $-31$ & $205$ & $-17$ & $220$ & $-68$ & $169$ \\
$32$ & $12 $ & $283$ & $-40$ & $232$ & $-18$ & $254$ & $-72$ & $200$ \\
\hline
\end{tabular}
\end{center}

{\bf Table 6.} {\small The total binding energies differences and binding energies themselves 
(in $Mev$) for the
 anti-flavored states, $SK4$-variant (first $4$ columns), and $SK6$ variant 
(last $4$ columns). $F_D/F_\pi=1.5, \; F_B/F_\pi=2.$}\\

For anti-charm and anti-beauty there is considerable difference between $SK4$ and $SK6$ variants
({\bf Tables 6,7}).
The mass of $\Theta_c$ hyperon in the $SK4$ model equals 
to $2700\,Mev$ and of $\Theta_b$ - to $5880 \,Mev$, both well below threshold for strong decay. For 
the $SK6$ variant these masses are by $40$ and $100\,Mev$ greater, but also below threshold. 
The $SK6$ variant is less attractive than $SK4$ variant, mainly due to the fact that the antiflavor 
excitation energies for $B=1$ in the $SK6$ variant are smaller than for $B\geq 2$, and this leads to 
repulsion for $B>1$, in comparison with the more familiar $SK4$ model. Considerable decrease of
binding energies for large $B$, greater than $B \sim 20$, may be connected with fact that the 
rational map approximation becomes to be unrealistic for such big baryon numbers. The beauty decay 
constant $F_b$ is not measured yet, and this introduces additional uncertainty in our predictions. 
Probably, the value $F_b/F_\pi =1.8$ is the best one for description of the $\Lambda_b$ mass.

In {\bf Table 7} we present the binding energies of hypernuclei with anti-charm and
anti-beauty quantum numbers for rescaled $SK4$ and $SK6$ variants of the model.

Several peculiarities should be emphasized. The binding energies for rescaled variants are
in general smaller than those for original variants ({\bf Table 6}), mainly due to the decrease
of excitation energies for the $B=1$ configuration (by $\sim 100 \,Mev$ for the anti-charm and 
$110\,Mev$ for anti-beauty). For greater $B$-numbers this decrease is smaller. Since, however,
rescaled or nuclear variant is valid for large enough baryon numbers, the binding energies can 
be greater than the values given in both {\bf Tables 6,  7}, at least for $B$-numbers 
greater than $\sim 10$. This is similar to the situation with strangeness quantum number 
({\bf Table 5} and its discussion).
\begin{center}
\begin{tabular}{|l|l|l|l|l||l|l|l|l|l|l|l|l|}
\hline
$B$ & $\Delta \epsilon_{\bar{c}}^{SK_4^*}$&
$\epsilon_{\bar{c}}$& $\Delta \epsilon_{\bar{b}}^{SK_4^*}$& $\epsilon_{\bar{b}}$ &
$\Delta \epsilon_{\bar{c}}^{SK_6^*}$&
$\epsilon_{\bar{c}}$& $\Delta \epsilon_{\bar{b}}^{SK_6^*}$& $\epsilon_{\bar{b}}$ \\
\hline
$2 $ & $36 $ & $36 $ & $54 $ & $54 $ & $-5 $ & $-5 $ & $-30$ & $-30$ \\
$3 $ & $24 $ & $32 $ & $35 $ & $43 $ & $-27$ & $-19$ & $-59$ & $-51$ \\
$4 $ & $19 $ & $48 $ & $44 $ & $72 $ & $-26$ & $2  $ & $-45$ & $-16$ \\
$5 $ & $27 $ & $54 $ & $39 $ & $66 $ & $-22$ & $5  $ & $-50$ & $-23$ \\
$6 $ & $18 $ & $50 $ & $31 $ & $63 $ & $-27$ & $5  $ & $-52$ & $-20$ \\
$7 $ & $30 $ & $69 $ & $46 $ & $84 $ & $-17$ & $22 $ & $-38$ & $1  $ \\
$8 $ & $19 $ & $75 $ & $27 $ & $84 $ & $-24$ & $32 $ & $-49$ & $7  $ \\
$9 $ & $21 $ & $78 $ & $23 $ & $80 $ & $-21$ & $36 $ & $-49$ & $8  $ \\
$10$ & $15 $ & $80 $ & $17 $ & $82 $ & $-25$ & $40 $ & $-52$ & $13 $ \\
$11$ & $17 $ & $91 $ & $13 $ & $88 $ & $-22$ & $52 $ & $-52$ & $23 $ \\
$12$ & $12 $ & $104$ & $9  $ & $101$ & $-25$ & $67 $ & $-53$ & $39 $ \\
$16$ & $3  $ & $131$ & $-12$ & $115$ & $-28$ & $100$ & $-61$ & $66 $ \\
$17$ & $6  $ & $136$ & $-10$ & $120$ & $-26$ & $104$ & $-60$ & $70 $ \\
$20$ & $-4 $ & $156$ & $-30$ & $131$ & $-31$ & $130$ & $-68$ & $93 $ \\
$24$ & $-10$ & $188$ & $-43$ & $155$ & $-33$ & $166$ & $-73$ & $125$ \\
$28$ & $-17$ & $220$ & $-57$ & $179$ & $-35$ & $202$ & $-78$ & $158$ \\
$32$ & $-21$ & $251$ & $-67$ & $205$ & $-37$ & $235$ & $-82$ & $190$ \\
\hline
\end{tabular}
\end{center}
{\bf Table 7.} {\small Same as in {\bf Table 6}, for rescaled $SK4$ and $SK6$ variants of the model.} \\
\section{Conclusions}
The excitation energies of antiflavors are estimated within the bound state version of
the chiral soliton model in two different variants of the model, $SK4$ and $SK6$, and for
two values of the model parameter ($e$ or $e'$, see {\bf Tables 3, 4}). The bounds for expected 
binding energies of hypernuclei are obtained in this way. These bounds are wide - variations of 
the total binding energy for the $SK4$ and $SK6$  models can reach $40-50\,Mev$.
The difference between original (baryon) variant and rescaled (nuclear) variant is greater for 
strangeness and smaller for anti-charm and anti-beauty, where
it is not greater than $\sim 20-30\; Mev$ for baryon numbers between $3$ and $\sim 20$. 
If the logic is correct, that rescaled or nuclear variant of the model should be applied for
large enough $B$-numbers, beginning with $B\sim 10$, then we should expect the existence of
weakly decaying hypernuclei with anti-charm and anti-beauty.

The properties of multiskyrmion configurations, necessary for these numerical estimates,
have been calculated within the rational map approximation \cite{hms} which provides remarkable 
scaling laws
 for the excitation energies of heavy antiflavors. This scaling property of heavy flavors
(antiflavors) 
excitation energies, noted previously \cite{kz,vk1}, and confirmed in present paper 
by numerical 
calculations, is fulfilled with good accuracy.
 The relative role of the quantum 
correction of the order $\sim 1/N_c$ (hyperfine splitting) decreases with increasing baryon number
like $1/B$, therefore, besides $1/N_C$ expansion widely used and discussed in the literature,
the $1/B$ expansion and arguments can be used to justify the chiral soliton approach at large
enough values of baryon number. 

Positive strangeness nuclear states are mostly bound relative to the decay into $\Theta^+$ and nuclear
fragments, so, one can speak about $\Theta^+$ hypernuclei \cite{mill,clm}. The particular value of 
binding energy
 depends on the variant of the model, and is greater for the original $SK4$ variant 
({\bf Table 5}). 
The 
existence of deeply bound states is not excluded by our treatment, although in most cases the 
energy
 of the state is sufficient for the strong decay into kaon and residual nucleus or nuclear 
fragments.

The binding energies of the ground states of hypernuclei with heavy antiflavors ($\bar{c}$ or 
$\bar{b}$) shown in {\bf Tables 6,7} are more stable relative to variations of the model 
parameters ($e$ or $e'$), but more sensitive to the model itself. Similar to the case of 
antistrangeness, the binding energies for the $SK6$ 
variant of the model are systematically smaller than for the $SK4$ variant.

Within our approach it is possible to obtain the spectra of excited states - with greater values of 
isospin and angular momentum. The energy scale in the first case is given by $1/\Theta_I$, in 
the second - by $1/\Theta_J$, which is much smaller for large baryon numbers. Since for $B=1$ 
$1/\Theta_I=1/\Theta_J \simeq 180\,Mev$, (see {\bf Table 1}) it seems difficult to obtain within 
chiral soliton approach such small spacing between ground state and excited levels as derived, 
e.g. in \cite{kimm2} within the quark models.

Although we performed considerable numerical work, we feel that further 
refinements, improvements as well as more precise calculations are necessary.
For example, possible contributions of the order $1/N_c$ to the
flavor excitation energies mentioned e.g. in \cite{wkleb} might change our conclusions.
When calculations for the present paper have been finished, we became aware of papers \cite{clm} 
and \cite{ztl}, where the possibility
of existence of anti-strange $\Theta$ hypernuclei is discussed within a framework of more 
conventional approaches.
 Results obtained in \cite{clm} and \cite{ztl} qualitatively agree with ours.

We thank V.A.Matveev and V.A.Rubakov for discussions and remarks. V.B.K. is indebted to
Ya.I.Azimov and I.I.Strakovsky for useful E-mail conversations, and to M.Karliner, H.Walliser and 
H.Weigel for numerous valuable discussions.
 The results of present paper have been reported 
in parts at the Conference QFTHEP, Peterhof, Russia, 19-25 June 2004 and Symposium of 
London Mathematical Society, Durham, UK, 2-12 August 2004.
\section{Appendix. Comparison of flavor and antiflavor excitation energy difference in rigid rotator
and bound state models}
Here we show that the difference between flavor and antiflavor excitation energies given by
(\ref{om}) coincides with the difference of $SU(3)$ rotation energies between exotic and 
nonexotic multiplets within rigid rotator approach, in the leading in $N_c$ approximation. 
The method used here is similar to that of \cite{vk} applied for arbitrary B-numbers and
$N_c=3$. Generalization for arbitrary $N_c$ and $N_F$ was made recently in \cite{jenkm}.
For nonexotic multiplets we have quantization condition \cite{g} $p+2q= N_cB$, and for odd 
$B$-numbers we take $p=1$, for even $B$, $p=0$.
The contribution to the $SU(3)$ rotation energy depending on "flavor" moment of inertia which is of
interest here equals to \cite{g}
\be
E^{rot}(SU_3) = {1\over 2\Theta_F}\bigl[C_2(SU_3)(p,q) - I_r(I_r+1) - N_c^2B^2/12\bigr]
\ee
with $C_2(SU_3) = (p^2+q^2+pq)/3 +p+q = [(p+2q)^2+3p^2]/12 +(p+2q)/2+p/2 $.
The "right" isospin for the lowest nonexotic states equals to $I_r=p/2=0$ for even $B$ 
(as for nuclei $^4He,\;^{12}C,\; ^{16}O$, etc.), and to $I_r=p/2=1/2$ for odd $B$ 
(as for isodoublets $^3H - ^3He,\; ^5He-^5Li$, etc.).

The lowest possible exotic $SU(3)$ irreps $(p,q)$ for each value of the baryon number 
$B$ are defined by
 relations:
$p'+2q'=N_cB+3m$; $m$ coincides with the number of additional quark-antiquark pairs for several
lowest values of $p'$. 

The difference of the $SU(3)$ rotation energies for exotic and nonexotic multiplets equals to
\be
\Delta E^{rot} = {1\over 2\Theta_{F,B}} \bigl[C_2(SU_3)' -C_2(SU_3) - I'_r(I'_r+1)+I_r(I_r+1)\bigr]
\ee
After simple transformations it can be written in the form:
\be
\Delta E^{rot} = {1\over 2\Theta_{F,B}} \bigl[\bigl(m(2N_cB+3m)+p'^2-p^2\bigr))/4 + 3m/2 + (p'-p)/2 +
(I_r-I'_r)(I_r+I'_r+1)\bigr].
\ee
If $m=1$, for lowest $SU(3)$ irreps $p'=1, \; q'=(N_cB+2)/2$ for even $B$, and 
$p'=0, \; q'=(N_cB+3)/2$ for odd $B$.
 We should keep in mind that the right isospin 
equals to $I'_r=(p'+1)/2=I_r+1$ for $B=2,\,4...$ and $I'_r=(p'+1)/2=I_r$ for $B=1,\,3,\,5...$.
For charm or beauty, due to large configuration mixing caused by large values
of $D$ or $B$-meson masses present in the lagrangian such
 lowest irreps often are not the main 
component of the mixed state (in this connection the papers \cite{bic} may be of interest), but 
for strangeness they are.

For even $B\;(m=1, p=1, p'=0)$ we have
\be
\Delta E^{rot} = {1\over 4\Theta_{F,B}} \bigl[N_cB + 2 \bigr].
\ee
For odd $B$ ($p=0, p'=1$) we obtain
\be
\Delta E^{rot} = {1\over 4\Theta_{F,B}} \bigl[ N_cB +3\bigr].
\ee
For $N_c=3$ and $B=1$ this coincides with well known expression for the mass difference between
andidecuplet and octet of baryons.

The leading in $N_c$ contribution is the same as given by (\ref{om}).
For arbitrary $m$ the leading contribution is $\Delta E^{rot} = mN_cB/(4\Theta_{F,B})$, for
any multiplets with the final values of $p'$ and $I_r$, also different from those we 
took here. It is worth noting that the correction to the leading contribution decreases not only
with increasing $N_c$, but also with increasing $B$ (recall that $\Theta_{F,B} \sim N_cB$).
Therefore, convergence of both methods is better for larger values of $B$.
\noindent
\baselineskip=10pt
\tenrm


\begin{thebibliography}{100}

\bibitem{japan}    T. Nakano et. al. [LEPS Collab.], Phys. Rev. Lett. {\bf 91}, 012002 (2003); 
 hep-ex/03010020;
 Barmin et al, [DIANA Collab.] Phys. Atom. Nucl. 66, 1715 (2003);hep-ex/0304040

\bibitem{step}  S. Stepanyan et al [CLAS Collab.] Phys. Rev. Lett. 91, 252001 (2003);
                    J. Barth  et al [SAPHIR Collab.], Phys. Lett. B572, 127 (2003);
              V. Kubarovsky and S. Stepanyan [CLAS Collab.] AIP Conf. Proc. 698, 543 (2003); 
	      hep-ex/0307088;
              A.E. Asratyan, A.G. Dolgolenko and M.A. Kubantsev, Phys. Atom. Nucl. 67, 682 (2004);
              V. Kubarovsky et al [CLAS Collab.] Phys. Rev. Lett. 92, 032001 (2004); 
              A. Airapetian et al [HERMES Collab.] Phys. Lett. B585, 213 (2004); 
              R. Togoo et al Proc. Mongolian Acad. Sc.  4, 2 (2003) ;
              A. Aleev et al [SVD Collab.] Yad. Fiz. (in press); hep-ex/0401024;
              M. Abdel-Bary et al [COSY-TOF Collab.] hep-ex/0403011;
	      S. Chekanov et al [ZEUS Collab.] Phys. Lett. B591, 7 (2004); hep-ex/0403051;
              M. Barbi [for ZEUS and H1 Collab.] hep-ex/0407006
\bibitem{kl}       M. Karliner and H.J. Lipkin, hep-ph/0307243; Phys. Lett. B575, 249 (2003); 
                hep-ph/0307343;
 hep-ph/0401072
\bibitem{jw}       R. Jaffe, F. Wilczek, hep-ph/0307341; Phys. Rev. Lett. 91, 232003 (2003)
; 
hep-ph/0312369

\bibitem{close}    F. Close, {\tenit The End of the Constituent Quark Model?}, Summary talk at
    the Baryon03 Conference,  hep-ph/0311087; AIP Conf. Proc. 717, 919 (2004); hep-ph/0411396

\bibitem{cld}      J.J. Dudek and F.E. Close, hep-ph/0311258;
                   F.E. Close and J.J. Dudek, hep-ph/0401192
\bibitem{kimm}  K. Maltman, hep-ph/04080144; hep-ph/0412328
\bibitem{pqw}  J. Ping, D. Qing, F. Wang and T. Goldman, hep-ph/0408176

\bibitem{alt}      C. Alt et al [NA49 Collab.] Phys. Rev. Lett. 92, 042003 (2004); hep-ex/0310014
\bibitem{fw}       H.G. Fischer and S. Wenig, hep-ex/0401014
; Eur. Phys. J. C37, 133 (2004)
\bibitem{akt}      A. Aktas et al [H1 Collab.] Phys. Lett. B588, 17 (2004); hep-ex/0403017

\bibitem{wa89} M.I. Adamovich et al [WA89 Collab.] hep-ex/0405042; U. Karshon [for ZEUS Collab.]
hep-ex/0407004
; I. Abt [for the HERA-B Collab.] hep-ex/0408048; B. Aubert [for the BABAR Collab.]
hep-ex/0408064
\bibitem{pochod} J. Pochodzalla, hep-ex/0406077
\bibitem{burkert} V.D. Burkert, R. De Vita, S. Niccolai and the CLAS Collaboration,hep-ex/0408019;
 K. Hicks, hep-ph/0408001; hep-ex/0412048; V.D. Burkert, nucl-ex/0412033
\bibitem{j}   R.L. Jaffe, SLAC-PUB-1774, Talk presented at the Topical Conf. on Baryon
              Resonances, Oxford, England, July 5-9, 1976; Phys. Rev. D15, 281 (1977)

\bibitem{hs} H. Hogaasen and P. Sorba, Nucl. Phys. B145, 119 (1978); D. Strottman, 
Phys. Rev. D20, 748 (1979)
\bibitem{ro}  C. Roiesnel,  Phys. Rev. D20, 1646 (1979)

\bibitem{manoh}  A.V. Manohar, Nucl. Phys. B248, 19 (1984); M. Chemtob,  
Nucl. Phys. B 256, 600 (1985).
\bibitem{km}  M. Karliner and M.P. Mattis, Phys. Rev. D34, 1991 (1986)

\bibitem{bieden}   L.C. Biedenharn and Y. Dothan, {\tenit Monopolar Harmonics in $SU_f(3)$
as Eigenstates of the Skyrme-Witten Model for Baryons}, in {\tenit From $SU(3)$ to Gravity}
(Ne'eman Festschrift) (Cambridge Univ. Press 1986).

\bibitem{mic}  M. Praszalowicz, {\tenit SU(3) Skyrmion} TPJU-5-87, Talk at the Cracow Workshop on 
Skyrmions
 and Anomalies, Mogilany, Poland, Febr. 20-24, 1987.

\bibitem{vk}  V.B. Kopeliovich, {\tenit On exotic systems of baryons in chiral soliton models}. 
NORDITA Preprint 90/55 NP (1990);  Phys. Lett. B 259, 234 (1991)

\bibitem{hans} H. Walliser, Nucl. Phys. A 548, 649 (1992);
{\tenit An extension of the standard Skyrme model} Proc. of the Workshop 
{\tenit Baryons as Skyrme Solitons}, p. 247,
 ed. G. Holzwarth, World Scientific (1992).

\bibitem{dpp} D. Diakonov, V. Petrov, and M. Polyakov,
 Z. Phys. A 359, 305 (1997)

\bibitem{herbert}  H. Weigel, Eur. Phys. J. A2, 391 (1998)

\bibitem{ja2}  R.L. Jaffe, hep-ph/0401178, v2; hep-ph/0405268
\bibitem{kss} V.B. Kopeliovich, B. Schwesinger and B.E. Stern,
 Nucl. Phys. A549, 485 (1992)
\bibitem{dpp2} D. Diakonov, V. Petrov and M. Polyakov, hep-ph/0404212
\bibitem{tc} T. Cohen, Phys. Lett. B581, 175 (2004); hep-ph/0309111

\bibitem{ikor} N. Itzhaki, I.R. Klebanov, P. Ouyang and L. Rastelli, Nucl. Phys. B684,264 (2004);
 I. Klebanov and P. Ouyang, hep-ph/0408251

\bibitem{wk}  H. Walliser and V.B. Kopeliovich, JETP 97, 433 (2003); hep-ph/0304058
\bibitem{bfk}   D. Borisyuk, M. Faber and A. Kobushkin, hep-ph/0307370; Ukr. J. Phys. 49, 944 (2004)
\bibitem{mic2}  M. Praszalowicz, Phys. Lett. B575, 234 (2003); hep-ph/0308114
\bibitem{wm1} B. Wu and B-Q. Ma, hep-ph/0312041; hep-ph/0408121
\bibitem{ekp}  J. Ellis, M. Karliner and M. Praszalowicz,
 JHEP 0405002 (2004); hep-ph/0401127
\bibitem{dpt}  G. Duplancic, H. Pasagic and J. Trampetic, hep-ph/0404193;
hep-ph/0405029; JHEP 0407, 027 (2004); B.J. Park, M. Rho and D.P. Min, hep-ph/0405246; 
H.K. Lee and H.Y. Park, hep-ph/0406051
\bibitem{w} H. Weigel, Eur. Phys. J. A21, 133 (2004); hep-ph/0404173; hep-ph/0405285
\bibitem{jm}  B.K. Jennings and K. Maltman, Phys. Rev. D69, 094020 (2004);
 V.B. Kopeliovich, Physics-Uspekhi 47, 309 (2004); 
D. Diakonov, hep-ph/0406043
; S.L. Zhu, hep-ph/0406204; M.Oka, hep-ph/0406211
; Ya.I. Azimov and
I.I. Strakovsky, hep-ph/0406312

\bibitem{l} H.J. Lipkin, Phys. Lett. B195, 484 (1987)

\bibitem{rs} D.O. Riska and N.N. Scoccola, Phys. Lett. B299, 338 (1993)

\bibitem{opm} Y. Oh, B.Y. Park and D.P. Min, Phys. Lett. B331, 362 (1994); 
Phys. Rev. D50, 3350 (1994)

\bibitem{ksr} V.B. Kopeliovich and M.S. Sriram, Phys. Atom. Nucl. 63, 480 (2000);hep-ph/9809394
\bibitem{kz} V.B. Kopeliovich and W.J. Zakrzewski, Eur. Phys. J. C18, 369 (2000);
                 JETP Lett. 69, 721 (1999); hep-ph/9904386
\bibitem{wkleb}  K.M. Westerberg and I.R. Klebanov, Phys. Rev. D50, 5834 (1994);
Phys. Rev. D53, 2804 (1996)		 
	

\bibitem{ck} C.G. Callan and I.R. Klebanov, Nucl.Phys. B262, 365 (1985); N. Scoccola, H. Nadeau,
M. Nowak and M. Rho, Phys.Lett. B201, 425 (1988)	 
\bibitem{hms}   C.J. Houghton, N.S. Manton and P.M. Sutcliffe,  Nucl. Phys. B510, 507 (1998)
\bibitem{kss2} V. Kopeliovich, B. Schwesinger and B. Stern, JETP Lett. 62, 185 (1995)		 
\bibitem{vkj} V. Kopeliovich, hep-ph/0310071; Proc. of the Conf. {\tenit
 Electroproduction of
Strangeness on Nucleons and Nuclei}, Sendai, Japan, 15-18 June 2003, pp. 96-111, Ed. K. Maeda, 
H. Tamura, S.N. Nakamura, O. Hasimoto, World Scientific, 2004.
\bibitem{cheung} K. Cheung, hep-ph/0308176
; B. Wu and B-Q. Ma, Phys. Rev. D70, 034025 (2004)
\bibitem{hldcz} P.-Z. Huang et al, hep-ph/0401191; I. Stewart, M. Wessling and M. Wise, 
hep-ph/0402076
\bibitem{bic} P. Bicudo, hep-ph/0403295; P. Bicudo, G.M. Marques, Phys. Rev. D69, 011503 (2004)
\bibitem{ch} T-W. Chiu and T-H. Hsieh, hep-ph/0404007; H. Kim and S.H. Lee, hep-ph/0404170 
\bibitem{kimm2} K. Maltman, hep-ph/0408145; hep-ph/0412327
\bibitem{mill} G.A. Miller, nucl-th/0402099; Phys. Rev. C70, 022202 (2004)

\bibitem{skyrme} T.H.R. Skyrme,
 Proc. Roy. Soc. A 260, 127 (1961); Nucl. Phys. 31, 556 (1962).
\bibitem{witten}  E. Witten,
 Nucl. Phys. B223, 422, 433 (1983).
\bibitem{jack} A. Jackson et al, Phys. Lett. B154, 101 (1985); A. Jackson, A.D. Jackson and
V. Pasquier, Nucl. Phys. A432, 567 (1985)
\bibitem{lm} L. Marleau, Phys. Lett. B235, 141 (1990); Erratum: Phys. Lett. B244, 580 (1990);
Phys. Rev. D45, 1776 (1992); L. Marleau and J.-F. Rivard, Phys. Rev. D63, 036007 (2001)
\bibitem{fp}  I. Floratos and B.M.A.G. Piette, Phys. Rev. D64, 045009 (2001); J. Math. Phys.
42, 5580 (2001)
\bibitem{sw}  B. Schwesinger and H. Weigel, Phys. Lett. B267, 438 (1991); H. Weigel, Int.J.Mod.Phys.
A11, 2419 (1996)

\bibitem{g}  E. Guadagnini, Nucl. Phys. B236, 35 (1984)
\bibitem{bps} R.A. Battye and P.M. Sutcliffe, Rev. Math. Phys. 14, 29 (2002)
\bibitem{vk1}  V.B. Kopeliovich, JETP 93, 435 (2001)
; J.Phys. G28, 103 (2002); 
JETP Lett. 73, 587 (2001)
\bibitem{ash}  A.M. Shunderuk, Yad. Fiz. 67, 769 (2004) [Phys. Atom. Nucl. 67, 748 (2004)]

\bibitem{ksm} V.B. Kopeliovich, A.M. Shunderuk and G.K. Matushko, nucl-th/0404020
; Phys. Atom. Nucl.
(2005), in press.
\bibitem{vk3} V.B. Kopeliovich, JETP, 96, 782 (2003)

\bibitem{aps} S. Nussinov, hep-ph/0307357; R.A. Arndt, I.I. Strakovsky and R.L. Workman,
Phys. Rev. C68, 042201 (2003); R.L. Workman et al, nucl-th/0410110; W.R. Gibbs, nucl-th/0405024
\bibitem{mic3} M. Praszalowicz, Phys. Lett. B583, 96 (2004); hep-ph/0311230

\bibitem{io} B.L. Ioffe and A.G. Oganesian, hep-ph/0408152
\bibitem{jenkm} E. Jenkins and A. Manohar, hep-ph/0402024; Phys. rev. D70, 034023 (2004)
\bibitem{clm} D. Cabrera, Q.B. Li, V.K. Magas, E. Oset, and M.J. Vicente Vacas, nucl-th/0407007
\bibitem{ztl} X.H. Zhong, Y.N. Tan, L.Li and P.Z. Ning, nucl-th/0408046
\end{thebibliography}
\end{document}